\documentclass[aps, pre, reprint, twocolumn, groupedaddress, showpacs, floatfix]{revtex4-1}

\usepackage{dcolumn}
\usepackage{amsmath, amssymb}
\usepackage[english]{babel}
\usepackage[utf8]{inputenc}

\usepackage{xy}
\usepackage{hyperref,graphicx}
\usepackage{verbatim}
\usepackage{enumerate}
\usepackage{url}
\usepackage{bm}
\usepackage{times}
\usepackage{helvet}
\usepackage{tikz}
\usetikzlibrary{shapes.geometric,arrows}
\tikzstyle{arrow} = [thick,->,>=stealth]

\newcommand{\note}[1]{}

\usepackage{xcolor}
\definecolor{darkblue}{RGB}{0,0,140}
\hypersetup{colorlinks,%
citecolor=darkblue,%
filecolor=green,%
linkcolor=blue,%
urlcolor=blue,%
}

\newcommand{\PFone}{{\fontfamily{phv}\selectfont PF$_1$}}
\newcommand{\PFtwo}{{\fontfamily{phv}\selectfont PF$_2$}}

\newcommand{\HB}{{\fontfamily{phv}\selectfont HB}}
\newcommand{\HBustb}{{\fontfamily{phv}\selectfont HB$_+$}}
\newcommand{\HBstab}{{\fontfamily{phv}\selectfont HB$_-$}}
\newcommand{\HBustbstab}{{\fontfamily{phv}\selectfont HB$_{+-}$}}

\newcommand{\HBprime}{{\fontfamily{phv}\selectfont HB'}}
\newcommand{\HC}{{\fontfamily{phv}\selectfont HC}}
\newcommand{\HCustb}{{\fontfamily{phv}\selectfont HC$_+$}}
\newcommand{\HCstab}{{\fontfamily{phv}\selectfont HC$_-$}}

\newcommand{\HCprime}{{\fontfamily{phv}\selectfont HC'}}

\newcommand{\SNone}{{\fontfamily{phv}\selectfont SN$_1$}}
\newcommand{\SNtwo}{{\fontfamily{phv}\selectfont SN$_2$}}
\newcommand{\SNthree}{{\fontfamily{phv}\selectfont SN$_3$}}
\newcommand{\SNfour}{{\fontfamily{phv}\selectfont SN$_4$}}

\newcommand{\SNLCone}{{\fontfamily{phv}\selectfont SNLC$_1$}}
\newcommand{\SNLCtwo}{{\fontfamily{phv}\selectfont SNLC$_2$}}
\newcommand{\CP}{{\fontfamily{phv}\selectfont CP}}
\newcommand{\CPC}{{\fontfamily{phv}\selectfont CPC}}
\newcommand{\GH}{{\fontfamily{phv}\selectfont GH}}
\newcommand{\ZH}{{\fontfamily{phv}\selectfont ZH}}
\newcommand{\BT}{{\fontfamily{phv}\selectfont BT}}
\newcommand{\BTprime}{{\fontfamily{phv}\selectfont BT'}}
\newcommand{\SLH}{{\fontfamily{phv}\selectfont SLH}}

\newcommand{\stateQ}{{Q}}
\newcommand{\stateS}{{S}}
\newcommand{\stateQQ}{{QQ}}
\newcommand{\stateSS}{{SS}}
\newcommand{\stateQS}{{QS}}
\newcommand{\stateSQ}{{SQ}}
\newcommand{\stateQSo}{{QS$_\text{o}$}}
\newcommand{\stateSQo}{{SQ$_\text{o}$}}

\newcommand{\maxdim}{N}
\newcommand{\maxpop}{M}

\newcommand{\R}{\mathbb{R}}
\newcommand{\T}{\mathbb{T}}

\newcommand{\Z}{\mathbb{Z}}
\newcommand{\N}{\mathbb{N}}
\newcommand{\C}{\mathbb{C}}

\newcommand{\thetasig}{\theta_{\sigma,k}}

\newcommand{\half}{\frac{1}{2}}

\def\txtd{{\textnormal{d}}}

\newcommand{\twocol}[1]{#1}

\begin{document}

\preprint{AIP/123-QED}


\title{
Birth and destruction of collective oscillations in a network of two populations of coupled type 1 neurons}%
\author{Benjamin J\"{u}ttner${}^{a}$}%
\author{Christian Henriksen${}^{a}$}%
\author{Erik A.~Martens${}^{a,b,c}$}%
\email{eama@dtu.dk}

\affiliation{%
\twocol{\mbox}
{${}^a$Department of Applied Mathematics and Computer Science, Technical
University of Denmark, 2800 Kgs.~Lyngby, Denmark}\\
\twocol{\mbox}
{${}^b$Department of Biomedical Sciences, University of Copenhagen, Blegdamsvej
3, 2200 Copenhagen, Denmark}\\
{${}^c$Center for Translational Neurosciences, University of Copenhagen,
Blegdamsvej 3, 2200 Copenhagen, Denmark}\\
}
\date{\today}

\begin{abstract}%
We study the macroscopic dynamics of large networks of excitable type 1 neurons composed of two populations interacting with disparate but symmetric intra- and inter-population coupling strengths.
This nonuniform coupling scheme facilitates symmetric equilibria, where both populations display identical firing activity, characterized by either quiescent or spiking behavior; or asymmetric equilibria, where the firing activity of one population exhibits quiescent but the other exhibits spiking behavior. Oscillations in the firing rate are possible if neurons emit pulses with non-zero width, but are otherwise quenched.  Here, we explore how collective oscillations emerge for two statistically identical neuron populations in the limit of an infinite number of neurons. A detailed analysis reveals how collective oscillations are born and destroyed in various bifurcation scenarios and how they are organized around higher codimension bifurcation points. Since both symmetric and asymmetric equilibria display bistable behavior, a large configuration space with steady and oscillatory behavior is available. Switching between configurations of neural activity is relevant in functional processes such as working memory, and the onset of collective oscillations in motor control.
\end{abstract}

\maketitle

\section*{}{\bf
The Theta neuron model~\cite{Ermentrout1986} is the normal form for the saddle-node-on-invariant cycle bifurcation, i.e., it represents the dynamic behavior near the excitation threshold of type 1 neurons, and it is equivalent to the quadratic integrate-and-fire neuron~\cite{Latham2000,Hansel2001, Gerstner2014}.
These neuron models have attracted much interest based on recently developed dimensional reduction techniques~\cite{OttAntonsen2008,Montbrio2015}, allowing for an exact description of neuron ensembles in terms of macroscopic collective variables~\cite{Luke2013,Montbrio2015}, for reviews see also~\cite{BickMartens2020,Byrne2019}. Such neuron populations mimic densely connected neural masses in the brain. Collective oscillations arising in the brain are important for generating rhythms in the brain, e.g., for motor control~\cite{Marder2001} and breathing~\cite{Smith1991}. The combination of excitatory and inhibitory neurons is a known prerequisite for the generation of collective rhythms such as gamma rhythms~\cite{Buzsaki2012a}. In this study we pursue the mathematical question of how collective rhythms may arise in an even simpler model composed of two populations of (statistically) identical excitatory neurons with nonuniform coupling, and what their bifurcations are.
}

\section{Introduction}
The brain is a complex network of networks with hierarchical structure~\cite{Bullmore2009,Meunier2010}, thus organizing neurons into neural masses, communities with high connectivity, structures which may interact with one another~\cite{Harris2005,Meunier2010} to solve cognitive functions~\cite{Lynn2019} by displaying different individual collective dynamic behaviors.  A prominent collective behavior observed in the brain occurs when a group of neurons synchronizes and oscillates in unison~\cite{Glass2001,Uhlhaas2006}. Synchrony has been associated with solving functional tasks including memory~\cite{Fell2011}, computational functions~\cite{Fries2009}, cognition~\cite{Wang2010}, attention~\cite{Singer1995,Fries2009}, processing and routing of information~\cite{Fries2005,Rabinovich2012,Kirst2016,DeschleDaffertshoferBattagliaMartens2019}, control of gait and motion~\cite{Marder2001}, or breathing~\cite{Smith1991}.

Neural masses with densely connected neurons are interconnected and form networks of modular structure. An important functional aspect in such networks are situations under which each population may assume different collective dynamic behaviors, such as low or high synchrony, or low and high firing activity. Thus, a network of oscillator populations may exhibit a large configuration space with different synchronization patterns, as is also exemplified by chimera states in Kuramoto oscillator networks, where one or several populations are synchronized and the other desynchronized~\cite{Abrams2008,MartensPanaggioAbrams2016,Martens2010bistable,Martens2016,DeschleDaffertshoferBattagliaMartens2019,Laing2009,Laing2012a,Scholl2016,PanaggioAbramsReview2015}. The dynamics of such networks with multi-population structure and their configurations has been explored in the context of neuroscience~\cite{Laing2016,Luke2013,Luke2014}, including memory recall~\cite{Schmidt2018}, information processing via self-induced stochastic resonance~\cite{YamakouHjorthMartens2020}, and deep brain stimulation~\cite{Weerasinghe2018}.

Many studies concern the modeling of neuronal processes at the microscopic scale of individual neurons. However, the number of neurons in the brain is enormous, and, consequently, mathematical models of the brain are very high dimensional, so that analyzing the collective dynamic behavior of large neuronal assemblies poses a prohibitive challenge; a coarse-grained description of the dynamics at the macroscopic level is desirable. Recently developed mathematical methods, based on the Ott-Antonsen~\cite{OttAntonsen2008} and Watanabe-Strogatz reductions~\cite{Watanabe1994,Laing2018} allow for an exact dimensional reduction, which applies to phase oscillator networks with sinusoidal coupling, including variants of the Kuramoto model, the theta neuron model and the equivalent quadratic integrate-and-fire neuron model. Unlike heuristic models~\cite{Wilson1972,Amari1977}, the  resulting model equations exactly describe the collective dynamics for each population, and -- connecting the microscopic to the macroscopic description -- accurately capture microscopic properties of the underlying system~\cite{Lin2020,BickMartens2020,Byrne2019}.

Collective oscillations in neural activity occur over a broad range of frequencies and across many brain regions~\cite{Buzsaki2012}. Prominent are gamma frequency oscillations, relevant in connection with cognitive tasks~\cite{Fries2007gamma}, neuronal diseases~\cite{Uhlhaas2006},  motor control~\cite{Marder2001} and breathing~\cite{Smith1991}. Such collective oscillations are known to occur in neuron networks with excitatory and inhibitory coupling~\cite{Keeley2017,Keeley2019,Segneri2020,Lin2020}. Network models with (statistically) identical neurons emitting infinitely 'sharp' signal pulses as represented by Dirac distributions do not permit collective oscillations~\cite{Devalle2017}; conversely, collective oscillatory behavior is possible when the pulse width is non-zero~\cite{Luke2013,So2013}.

We study a network composed of two populations of inhibitory type 1 neurons with non-uniform (but symmetric) coupling, interacting through pulses with non-zero width. We consider the dynamics in the continuum limit of infinitely many neurons, allowing us to use aforementioned dimensional reduction methods~\cite{OttAntonsen2008,BickMartens2020}. Rather than aiming at a high level of biophysical realism, we wish to elucidate how collective oscillations may get born and destroyed in a simple setup and to explore their related bifurcation scenarios. Even though the coupling is symmetric and neurons are statistically identical, the resulting dynamic behavior is surprisingly complicated.
The neuronal activity in each population may assume distinct levels, thus resulting in multistable configurations, in similarity to  synchronization patterns as those observed in chimera states~\cite{PanaggioAbramsReview2015,Abrams2008}, or (non-oscillatory) neural states reported for models of working memory~\cite{Schmidt2018}. In particular, one observes a rich structure of bifurcations producing collective limit cycle oscillations for which we provide a detailed bifurcation analysis.

The article is structured as follows. In Sec.~II, we introduce our model of two populations of Theta neurons and its equivalent form of quadratic integrate-and-fire neurons. We outline how an exact description of the macroscopic dynamics for populations of infinitely many neurons is obtained via the Ott-Antonsen method, and how firing rate equations for the equivalent QIF neurons are derived via a conformal mapping~\cite{Montbrio2015}. In Sec.~III, we summarize the known dynamical behavior for a single population, which represents a limiting case for two populations with vanishing inter-population coupling or uniform coupling. In Sec.~IV, we perform a detailed analysis by using numerical continuation methods via MatCont~\cite{Dhooge2008}, and explain the various bifurcation scenarios that are possible. Finally, we sum our findings up and conclude with a discussion in Sec.~V.

\section{Model}%
\subsection{Network of Theta neurons}
We consider a model of $ M=2 $ populations of $ N $ interacting Theta neurons, where the phase $ \theta_{\sigma,k} \in \T :=\R/2\pi\Z $ of the $ k $th neuron belonging to population $ \sigma=1,2 $ evolves according to
\begin{align}\label{eq:thetamodel}
 \dot{\theta}_{\sigma,k}:=\frac{\txtd \theta_{\sigma,k}}{\txtd t} & =
 1-\cos{\theta_{\sigma,k}} + (1+\cos{\theta_{\sigma,k}} )(\eta_{\sigma,k} + I_\sigma)
\end{align}
with  excitability $ \eta_{\sigma,k} $ of oscillator $ k $ in population $ \sigma $ sampled from a Lorentzian distribution $ g_\sigma(\eta) $ with mode $ \hat{\eta}_\sigma $ and width $ \Delta_\sigma $.
The Theta neuron \eqref{eq:thetamodel} is the normal form of the saddle-node-on-invariant-circle (SNIC) (or saddle-node-infinite period) bifurcation~\cite{Ermentrout2008} and is a canonical type 1 neuron~\cite{Ermentrout1986}. The dynamics are as follows. For $\eta_{\sigma,k}+I_\sigma<0$, a stable and unstable fixed point occur on the phase circle $\T$; for $\eta_{\sigma,k}+I_\sigma=0$, these fixed points coalesce in a saddle-node bifurcation; for $\eta_{\sigma,k}+I_\sigma>0$, the flow on the circle results in a cyclic/periodic motion. If $ \eta_{\sigma,k}+I_\sigma < 0 $, the Theta neuron is said to be \textit{excitable}: in the absence of perturbations, the phase relaxes to the stable fixed point on the phase circle $\T$; however, a perturbation may lead to a single spike (at $ \thetasig=\pi $) before returning to the stable fixed point. This could happen in at least two ways: a perturbation of the phase across the unstable fixed point (constituting a threshold) is possible, if one considers that the Theta model derives from a higher dimensional model~\cite{Ermentrout1986} so that the circle is embedded in a higher dimensional space; alternatively, a very short-lived (time scale of a single cycle) increase in $I_\sigma$ momentarily pushes the system across the bifurcation threshold  $\eta_{\sigma,k}+I_\sigma=0$. If $ \eta_{\sigma,k}+I_\sigma>0 $, the neuron is \textit{firing} (or excited), i.e., it spikes periodically.

\begin{figure}[htp!]
  \centering
  \includegraphics[width=0.9\columnwidth]{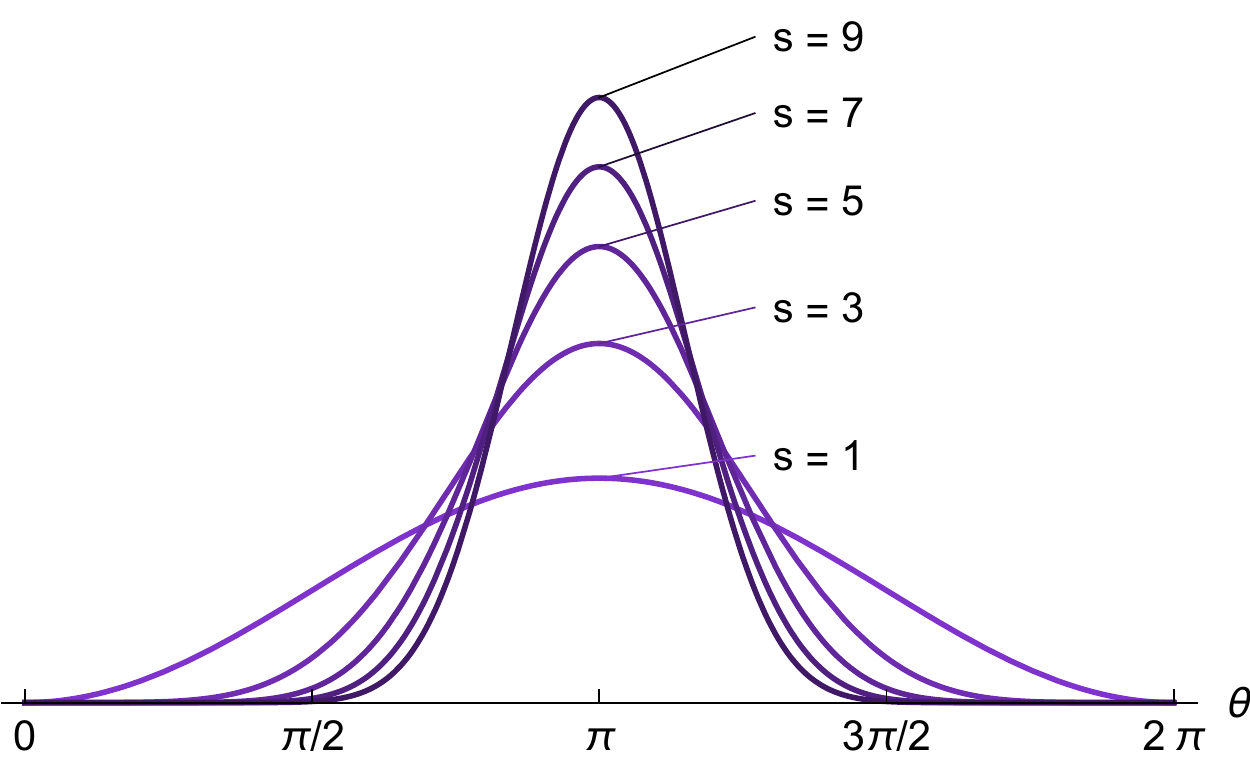}
  \caption{\label{fig:pulseshape} Pulse shape for varying pulse shape parameter $s$. The pulse converges to Dirac delta as $s\rightarrow \infty$.
  }
\end{figure}

The input current may result from a variety of interactions, for an overview see~\cite{BickMartens2020,Devalle2017}. Here, we assume that the input current is given by
\begin{align}\label{eq:coupling}
 I_\sigma &= \sum_{\tau=1}^\maxpop \frac{\kappa_{\sigma\tau}}{\maxdim}\sum_{l=1}^{\maxdim} P_s(\theta_{\tau,l})
\end{align}
where adjacent neurons interact via pulses, which we choose to be
\begin{align}\label{eq:AriaratnamStrogatzPulseShape}
P_s(\theta) = a_s (1-\cos{\theta})^s,\
\end{align}
originally adopted by Ariaratnam and Strogatz~\cite{ariaratnam2001}, with shape parameter $ s\in \N $, see also Fig.~\ref{fig:pulseshape}, and coupling strengths $\kappa_{\sigma\tau}$ between populations $\sigma$ and $\tau$.
The normalization constant $ a_s = 2^s (s!)^2/(2s)!$ is defined so that $ \int_0^{2\pi} P_s(\theta)d\theta = 2\pi $.

The case of $ M=2 $ populations results in eight parameters (excluding the pulse shape parameter $ s $). To reduce the problem to a manageable number of parameters, we make the following assumptions:
(i) the oscillator properties in populations $\sigma=1,2$ are  statistically identical so that $ \hat{\eta}_1=\hat{\eta}_2 =:\hat{\eta} $ and $ \Delta_1=\Delta_2 =:\Delta $; and
(ii) the coupling is symmetric with respect to identical intra- and inter-coupling strengths, i.e., $ \kappa_{11}=\kappa_{22}=:\kappa $. and $ \kappa_{12}=\kappa_{21}=:a \kappa $.
Unless stated otherwise, we keep $ (\hat{\eta},\Delta,s) $ fixed and consider $ (\kappa,a) $ the main bifurcation parameters.

\subsection{Network of Quadratic Integrate and Fire neurons}
An equivalent description of the Theta neuron is the quadratic integrate-and-fire (QIF) neuron via the transformation into the membrane potential $V(\theta)=\tan{(\theta/2)} \in(-\infty,\infty)$. The model equations then become~\cite{BickMartens2020}
\begin{align}
 \frac{\txtd}{\txtd t } V_{\sigma,k} := \dot{V}_{\sigma,k} &= V_{\sigma,k}^2 + \eta_{\sigma,k} + I_\sigma,\
\end{align}
where $V_{\sigma,k}:=V(\theta_{\sigma,k})$.
In this formulation, the neuron fires (emits a spike) when the voltage reaches $V_k(t^-)=\infty$ (in finite time). It is then reset to $V_k(t^+)=-\infty$.
QIF neurons have been widely used in neuroscientific modeling; see~\cite{Ermentrout2010,Gerstner2014} for a general introduction and~\cite{Hansel2001,Brunel2003} for a few examples of applications of QIF neurons.

\subsection{Exact macroscopic description for the limit of infinitely many neurons}%
We consider  \eqref{eq:thetamodel} in the limit $ N\rightarrow \infty $,
which allows us to express the ensemble dynamics
in terms of a continuous neuron density $\rho_\sigma(\theta,\eta,t)$ governed by the continuity equation
\begin{align}
    \frac{\partial }{\partial t}\rho_\sigma + \frac{\partial}{\partial \theta}({f_\sigma} \rho_\sigma) &=0\
\end{align}
where
\begin{align}
  f_\sigma &= 1-\cos\theta + (1+\cos\theta) \\ \nonumber
  &\times\left(  \eta + 
\sum_{\tau=1}^M \kappa_{\sigma\tau} \int_{-\infty}^{\infty}\int_0^{2\pi}P_s(\theta') \rho_\tau(\eta',\theta',t) \txtd\theta' \txtd\eta' \right)
.\
\end{align}
The Ott-Antonsen method~\cite{OttAntonsen2008,Luke2013} facilitates an exact reduction of the microscopic dynamics in \eqref{eq:thetamodel} to a low-dimensional description of the macroscopic dynamics in terms of the complex order parameter of each population,
\begin{align}
  Z_\sigma (t) &= R_\sigma(t)e^{-i\Phi_\sigma(t)} = \int_0^{2\pi}\int_{-\infty}^\infty e^{i\theta}\rho_\sigma(\theta,\eta,t)\,\txtd\eta\, \txtd\theta.\
\end{align}
The absolute value of the order parameter informs us of the level of phase synchronization of the neuron population: when $|Z_\sigma|\approx 0$, phases are spread over the circle $\T$, whereas $|Z_\sigma|\approx 1$ implies phase synchronization, i.e., phases are closely spread around the phase of the order parameter given by $\Phi_\sigma=-\arg{(Z_\sigma)}$. The collective dynamics of population $ \sigma = 1,2 $ is then given by~\cite{Luke2013,BickMartens2020}
\begin{align}\label{eq:OP_dynamics}
  \begin{split}
    \dot{Z}_\sigma &=
    -\frac{1}{2}\left[
    (\Delta_\sigma - i \hat{\eta}_\sigma - i I_\sigma^{(s)}) (1+Z_\sigma)^2
    + i(1-Z_\sigma)^2
    \right].\
  \end{split}
\end{align}
These equations are closed by the input current~\cite{Luke2013,BickMartens2020}
\begin{align}
 I_\sigma^{(s)} & = \sum_{\tau=1}^M \kappa_{\sigma\tau} P^{(s)}_\sigma,
 \end{align}
 with the average output from all other neurons in the network,
\begin{align}\label{eq:averageoutputneuron}
 P^{(s)}_\sigma &= a_s \left(C_0 + \sum_{q=1}^s C_q(Z_\sigma^q+\bar{Z}_\sigma^q)\right),\\
C_q &= \sum_{k=0}^s\sum_{m=0}^k \frac{s!(-1)^k\delta_{k-2m,q}}{2^km!(s-k)!(k-m)!}.\
\end{align}
For details on this reduction method and theory in general including applications in neuroscience, see~\cite{BickMartens2020}.

Two cases are of particular interest to us: pulse shape parameter $ s=1 $ and $ s=\infty $ (impulsive coupling) for which we have
\begin{align}\label{eq:P1Z}
 P^{(1)}_\sigma &= 1 - \half(Z_\sigma+\bar{Z}_\sigma),\
\end{align}
and
\begin{align}\label{eq:PinfZ}
 P^{(\infty)}_\sigma &= \frac{1-|Z_\sigma|^2}{(1+Z_\sigma)(1+\bar{Z}_\sigma)},\
\end{align}
respectively.

\subsection{Firing rate equations}

The model~\eqref{eq:OP_dynamics} has an equivalent formulation in terms of average firing rate $ r_\sigma $ and average membrane potential $ v_\sigma $ called the Firing Rate Equations (FRE)~\cite{Montbrio2015}. Indeed, changing variables via the (anti)conformal mapping
\begin{align}\label{eq:conformalmap}
  Z &=(1-\bar{W})/(1+\bar{W}) \quad\text{or}
   & W &= (1-\bar{Z})/(1+\bar{Z}),
\end{align}
gives
\begin{align}\label{eq:complexFRE}
 \dot{W}_\sigma&= \Delta_\sigma + i\hat{\eta}_\sigma - iW_\sigma^2 + i I_\sigma^{(s)}\
\end{align}
and
\begin{align}\label{eq:P1W}
 P^{(1)}_\sigma &= 1-\frac{1-|W_\sigma|^2}{(1+W_\sigma)(1+\bar{W_\sigma})}, \
\end{align}
\begin{align}\label{eq:PinfW}
 P^{(\infty)}_\sigma &= \half(W_\sigma+\bar{W}_\sigma).
\end{align}
Writing $ W_\sigma = \pi r_\sigma + i v_\sigma $, \eqref{eq:P1W} and \eqref{eq:PinfW} take the form
\begin{align}
  P^{(1)}_\sigma &= 2\dfrac{ \pi^2 r_\sigma^2 + \pi r_\sigma + v^2_\sigma}{(\pi r_\sigma+1)^2+v_\sigma^2},
  \quad
  P^{(\infty)}_\sigma = \pi r_\sigma, \
\end{align}
for $s=1$ and $s=\infty$, respectively.
Taking real and imaginary part of \eqref{eq:complexFRE} yields the firing rate equations
\begin{align}\label{eq:FREM1}
  \dot{r}_\sigma&=\frac{\Delta_\sigma}{\pi} +2 r_\sigma v_\sigma,
  \\
  \label{eq:FREM2}
  \dot{v}_\sigma&= v_\sigma^2  - \pi^2 r_\sigma^2 + \hat{\eta}_\sigma + I_\sigma^{(s)},\
\end{align}
where
\begin{align}
  I_\sigma^{(s)}&=
 \left\{
 \begin{array}{cl} 2\sum_{\tau=1}^M\kappa_{\sigma\tau}
 \dfrac{ \pi r_\tau^2 + \pi r_\tau + v_\tau^2}{(\pi r_\tau+1)^2+v_\tau^2} &,\quad s=1\\
   \pi \sum_{\tau=1}^M\kappa_{\sigma\tau} r_\tau\
 &, \quad s=\infty.\
 \end{array}
 \right.
 \
\end{align}
The microscopic and macroscopic description are related as follows. A single Theta neuron fires when its phase crosses $\theta=\pi$; accordingly, the average firing rate $r_\sigma(t)$ of the network at time $t$ is defined as the flux through $\theta=\pi$ (or equivalently, the flux at $v_\sigma=\infty$), see for instance~\cite{BickMartens2020}.

\section{Dynamic behavior of one population\label{sec:one_population}}
The dynamic behavior for the case of $M=1$ population has already been studied previously~\cite{Luke2013,So2013}. We briefly review the dynamics observed for this case as it is instructive for understanding  the dynamic and oscillatory behavior exhibited by $M=2$ populations.
For two parameter choices, the model equations~\eqref{eq:thetamodel} for $M=2$ effectively reduce to the dynamics of a single population, $M=1$. Recall that the intra- and inter-coupling strengths among the two populations are given via $\kappa_{11}=\kappa_{22}=\kappa$  and $\kappa_{12}=\kappa_{21}=\kappa a$.
Thus, when $a=1$, all neurons experience identical coupling strength so that the two populations act like a single population consisting of twice the number of neurons; on the other hand, when $a=0$, the two populations are decoupled so that each of the two populations in separation effectively corresponds to an $M=1$ system.  For brevity, we drop $\sigma$ in~\eqref{eq:thetamodel} and all related equations.

\begin{figure}[htp!]
\centering
\begin{tikzpicture}
  \draw (0, 0) node[inner sep=0] {\includegraphics[height=0.2\textheight]{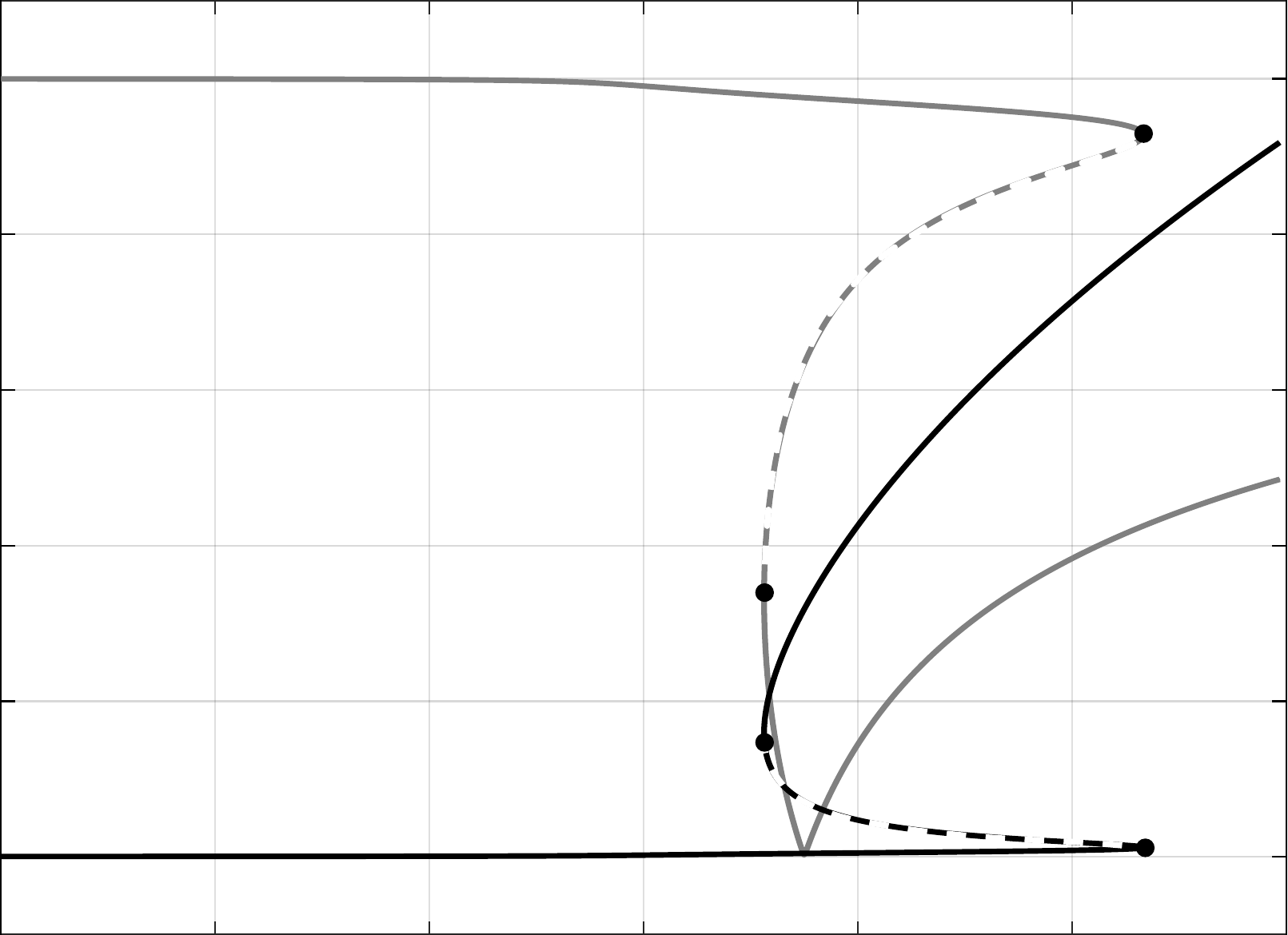}};
  \node[rotate=0] at (3.23, -2.55) {6};
  \node[rotate=0] at (-3.3, -2.55) {-6};
  \node[rotate=0] at (0, -2.55) {$\kappa$};
  \node[rotate=0] at (-3.5, -1.97) {0};
  \node[rotate=0] at (-3.5, 1.95) {1};
  \node[rotate=90, color=black] at (-3.75, 0) {$\text{min}_t (r_{}), \text{max}_t (r_{})$};
  \node[rotate=0] at (3.5, -1.97) {0};
  \node[rotate=0] at (3.5, 1.95) {1};
  \node[rotate=90, color=gray] at (3.75, 0) {$\text{min}_t (|Z_{}|), \text{max}_t (|Z_{}|)$};
  \node[rotate=0, color=gray] at (0.2, -0.6) {\SNone};
  \node[rotate=0] at (0.2, -1.4) {\SNone};
  \node[rotate=0, color=gray] at (2.9, 1.85) {\SNtwo};
  \node[rotate=0] at (2.9, -1.7) {\SNtwo};
  \node[rotate=0, color=black] at (-2.1, -1.8) {\footnotesize Quiescence (Q)};
  \node[rotate=45, color=black] at (2, 0.9) {\footnotesize Spiking (S)};
  \node[rotate=0] at (-4, 2.3) {a)};
\end{tikzpicture}\\
\begin{tikzpicture}
  \draw (0, 0) node[inner sep=0] {\includegraphics[height=0.2\textheight]{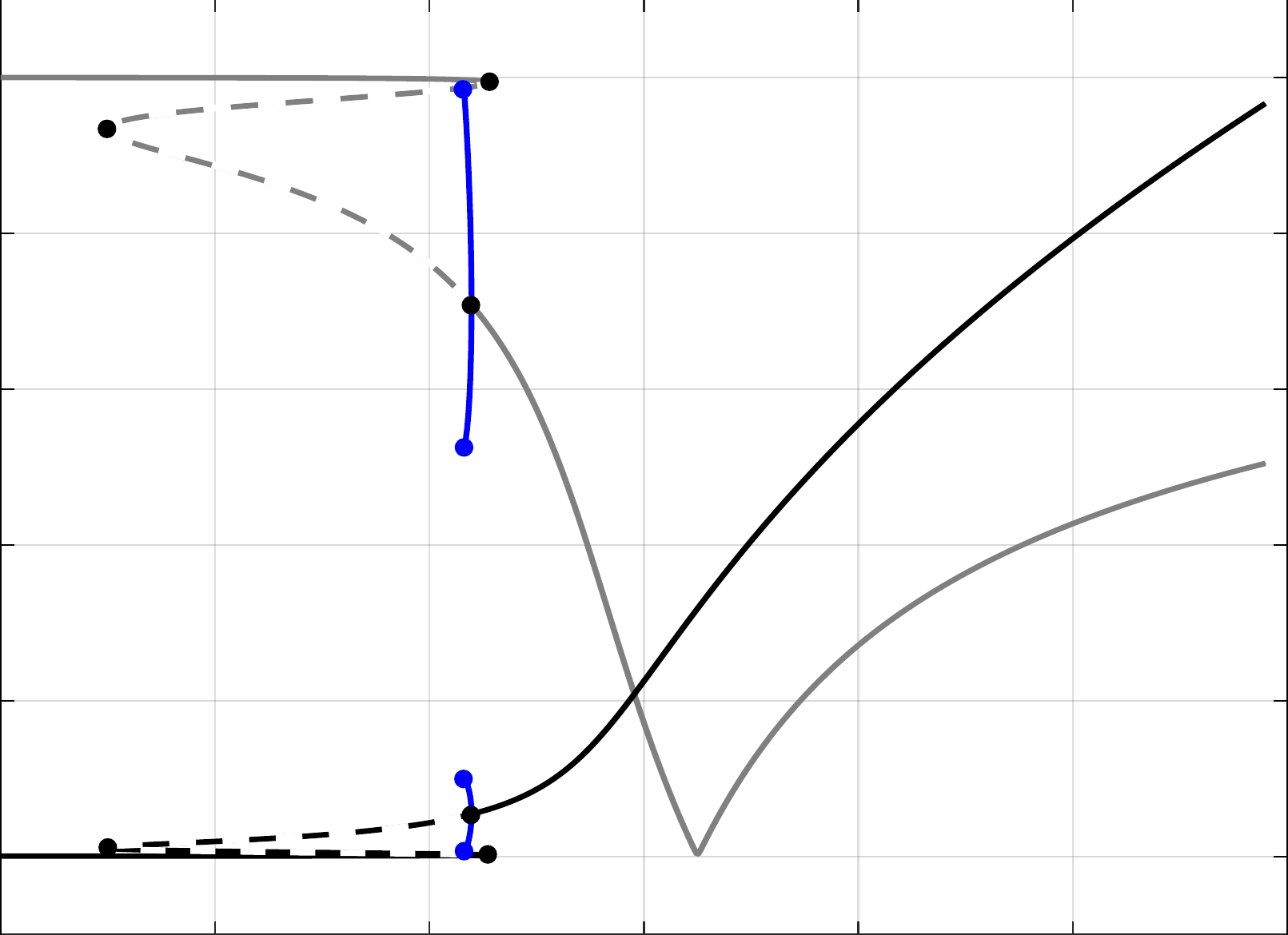}};
  \node[rotate=0] at (3.23, -2.55) {6};
  \node[rotate=0] at (-3.3, -2.55) {-6};
  \node[rotate=0] at (0, -2.55) {$\kappa$};
  \node[rotate=0] at (-3.5, -1.97) {0};
  \node[rotate=0] at (-3.5, 1.95) {1};
  \node[rotate=90, color=black] at (-3.75, 0) {$\text{min}_t (r_{}), \text{max}_t (r_{})$};
  \node[rotate=0] at (3.5, -1.97) {0};
  \node[rotate=0] at (3.5, 1.95) {1};
  \node[rotate=90, color=gray] at (3.75, 0) {$\text{min}_t (|Z_{}|), \text{max}_t (|Z_{}|)$};
  \node[rotate=0, color=gray] at (-2.95, 1.6) {\SNone};
  \node[rotate=0, color=gray] at (-0.3, 2.1) {\SNtwo};
  \node[rotate=0, color=gray] at (-0.4, 0.9) {\HB};
  \node[rotate=0, color=gray] at (-1.3, 0.1) {\HC};
  \node[rotate=0] at (-2.55, -1.7) {\SNone};
  \node[rotate=0] at (-0.6, -2.2) {\SNtwo};
  \node[rotate=0] at (-0.5, -1.1) {\HB};\draw (-0.5,-1.275) -- (-0.825,-1.7) ;
  \node[rotate=0] at (-1.2, -1.4) {\HC};
  \node[rotate=0, color=black] at (-2.3, -0.8) {\footnotesize Quiescence (Q)};
  \draw[arrow] (-3.,-1.) -- (-3,-1.9);
  \node[rotate=40, color=black] at (2, 1.25) {\footnotesize Spiking (S)};
  \node[rotate=0] at (-4, 2.3) {b)};
\end{tikzpicture}
  \caption{\label{fig:bifdiag_M1}
  Bifurcation diagrams in $\kappa$ for $M=1$ population of Theta neurons. We display solution branches by reporting maxima and minima in the firing rate $r$ (black) and synchrony level $|Z|$ (gray), respectively. Stable and unstable branches of equilibria have coinciding minima/maxima and are shown solid and dashed, respectively; minima/maxima corresponding to limit cycle behavior are highlighted in blue.  Bifurcations that may occur are: saddle node (\SNone, \SNtwo), Hopf (\HB)~ and homoclinic (\HC). Fixed parameters are $\Delta = 0.01 $ and pulse shape parameter $ s = 1 $, while $ \hat{\eta} = -0.5$ and $ \hat{\eta} = 0.5$ in panels a) and b), respectively.
  }
\end{figure}
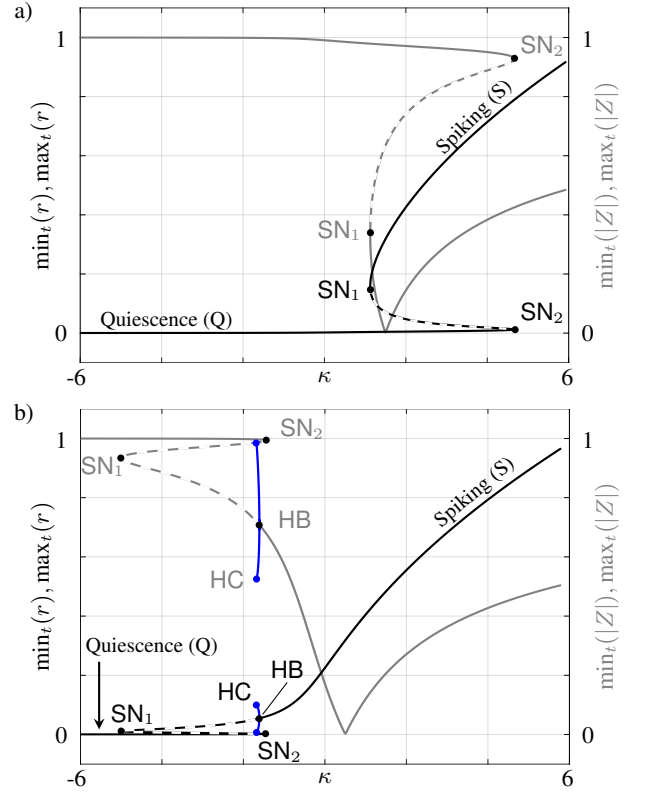

The bifurcation diagrams in Fig.~\ref{fig:bifdiag_M1} report
minima and maxima for the firing rate $r$ while varying coupling strength $\kappa$ with parameter values $s=1,\Delta=0.01$ fixed, and $\hat{\eta}=-0.5$ or $\hat{\eta}=0.5$ in panels a) and b), respectively.
Solution branches sometimes appear very close to each other for the firing rate $r$, therefore it is instructive to also report the magnitude of the order parameter, $|Z|$, which is related to the firing rate $r$ via the (anti)conformal mapping~\eqref{eq:conformalmap}.
Equilibria and local bifurcations (saddle-node, Hopf) can be computed analytically from~\eqref{eq:FREM1} and~\eqref{eq:FREM2}; limit cycles and other bifurcations were computed and continued numerically using Matlab and MatCont software~\cite{Dhooge2008}, see also Appendix~\ref{app:meth}.

We first consider the case of excitable neurons ($\hat{\eta}<0$) in Fig.~\ref{fig:bifdiag_M1}a). For the parameters considered and $\kappa \lessapprox 0$, we observe a set of stable equilibria (stable nodes) with $|Z|\approx1$; the related microscopic states are non-oscillatory, i.e., most of the neurons are \emph{quiescent} (Q), and so their spiking activity is negligible, $r\approx 0$. This branch of equilibria may undergo two saddle-node bifurcations (\SNone\, and \SNtwo) which are connected by a branch of saddles. Equilibria to the right of \SNone\, (larger $\kappa$) are stable spirals and correspond to \emph{spiking} neurons (S) with larger firing rate $r > 0$. As the  coupling strength $\kappa$ increases, higher levels of synchrony, eventually getting close to $|Z|=1$, may be achieved.

For the case of spiking (firing) neurons ($\hat{\eta}>0$), the bifurcation diagram in Fig.~\ref{fig:bifdiag_M1}b) reveals a similar bifurcation structure with two saddle-node bifurcations. However, for certain values of $\hat{\eta}$, an even more complicated bifurcation scenario is possible along the branch to the right of \SNone: a supercritical Hopf bifurcation (\HB) gives birth to limit cycles which ultimately are destroyed in a homoclinic bifurcation (\HC).
In between the values of $\hat{\eta}=-0.5$ and $\hat{\eta}=0.5$ shown in Fig.~\ref{fig:bifdiag_M1}, two distinct bifurcations of codimension 2 occur: (i) \SNone\ and \SNtwo\ merge in a cusp point, and (ii) the bifurcation curves \SNone, \HB\ and \HC\ meet in a Bogdanov-Takens point. The scenario in which limit cycles occur is characteristic for spiking neurons ($\hat{\eta}>0$) with inhibitory coupling ($\kappa<0$), as can be shown by further bifurcation analysis. For further details on these bifurcation structures, see~\cite{Luke2013,So2013}.

Importantly, we note that collective oscillations emerging in the Hopf bifurcation
\HB\ cease to exist in the limit of pulses defined by \eqref{eq:AriaratnamStrogatzPulseShape} with zero width obtained in the limit of $s\rightarrow \infty$. While this was already noted in recent studies~\cite{Ratas2016,Devalle2017} we briefly outline a derivation of this fact in  Appendix~\ref{app:collective_oscillations}. Further investigations of ours show that Hopf bifurcations continue to exist for a large range of values of the pulse shape parameter, $s$. Our observations suggest that the Hopf bifurcations giving birth to oscillations only vanish in the limit of $s\rightarrow\infty$, prompting a degeneracy for this limit.
The case of infinitely narrow pulses, $ s = \infty $, provides the advantage that the fixed point conditions resulting from the corresponding FRE can be solved in closed form, enabling a simple mathematical analysis. However, since this case produces a degenerate bifurcation behavior where limit cycles are absent, we chose to fix $s=1$.

Between the pair of fold bifurcations (\SNone\ and \SNtwo) a parameter region of bistability arises, thus facilitating hysteretic behavior. This happens for excitable neurons, $\hat{\eta}<0$, with \emph{excitatory coupling}, $\kappa>0$, as well as for parameters corresponding to firing neurons, $\hat{\eta}>0$, with inhibitory coupling, $\kappa<0$. This bistable character of solutions observed for $M=1$ population translates to the case of $M=2$ populations, where each population may attain distinct stable configurations.

In the following,  we consider non-zero pulse width ($s=1$) and fix parameter values to $\hat{\eta}=-1$ (excitable neurons) and $\Delta=0.01$, while varying the intra-coupling strength, $\kappa$, and the inter-coupling strength, $a$.

\section{Analysis for two populations\label{sec:analysis_twopopulations}}

\subsection{Symmetric and asymmetric equilibria}
\begin{figure}[htp!]
   \centering
   \begin{tikzpicture}
    \draw(0,0) node[inner sep=0]{\includegraphics[width=0.45\columnwidth]{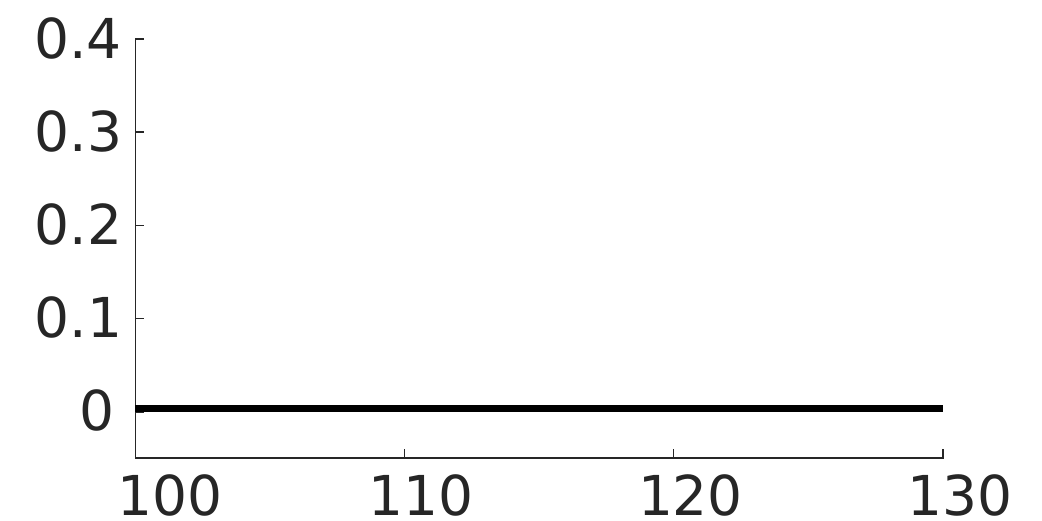}};
   \node[rotate=90, color=black] at (-2, 0) {$r_{1,2}$};
   \node[rotate=0] at (0, -1.25) {$t$};
   \node[rotate=0] at (0, 0) {\stateQQ};
   \node[rotate=0] at (-0.8, -0.35) {$r_1=r_2$};
    \node[rotate=0] at (-2, 1.) {a)};
   \end{tikzpicture}
   \begin{tikzpicture}
    \draw (0, 0) node[inner sep=0] {\includegraphics[width=0.45\columnwidth]{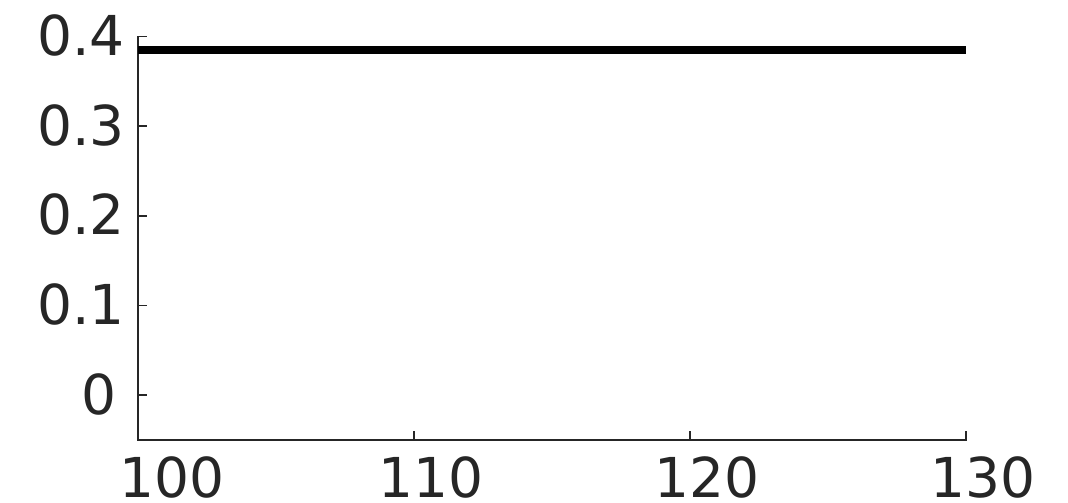}};
   \node[rotate=90, color=black] at (-2, 0) {$r_{1,2}$};
   \node[rotate=0] at (0, -1.25) {$t$};
   \node[rotate=0] at (0, 0) {\stateSS};
   \node[rotate=0] at (-0.8, 0.9) {$r_1=r_2$};
    \node[rotate=0] at (-2, 1.) {b)};
   \end{tikzpicture}
    \begin{tikzpicture}
    \draw (0, 0) node[inner sep=0] {\includegraphics[width=0.45\columnwidth]{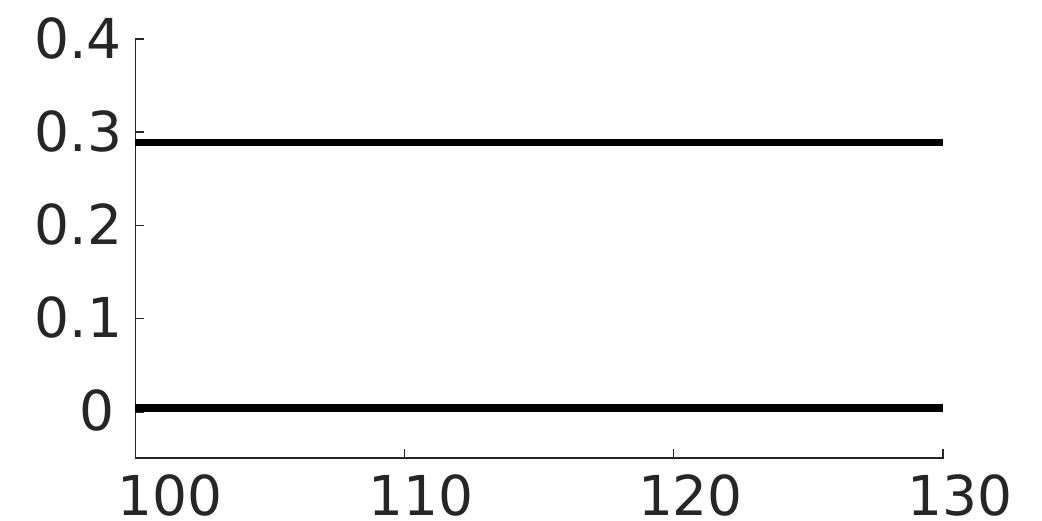}};
    \node[rotate=90, color=black] at (-2, 0) {$r_{1,2}$};
   \node[rotate=0] at (0, -1.25) {$t$};
   \node[rotate=0] at (0, -0.2) {\stateQS\, / \stateSQ};
   \node[rotate=0] at (-1.1, 0.6) {$r_{1,2}$};
   \node[rotate=0] at (-1.1, -0.4) {$r_{2,1}$};
    \node[rotate=0] at (-2, 1.) {c)};
    \end{tikzpicture}
    \begin{tikzpicture}
    \draw(0,0) node[inner sep=0]{\includegraphics[width=0.45\columnwidth]{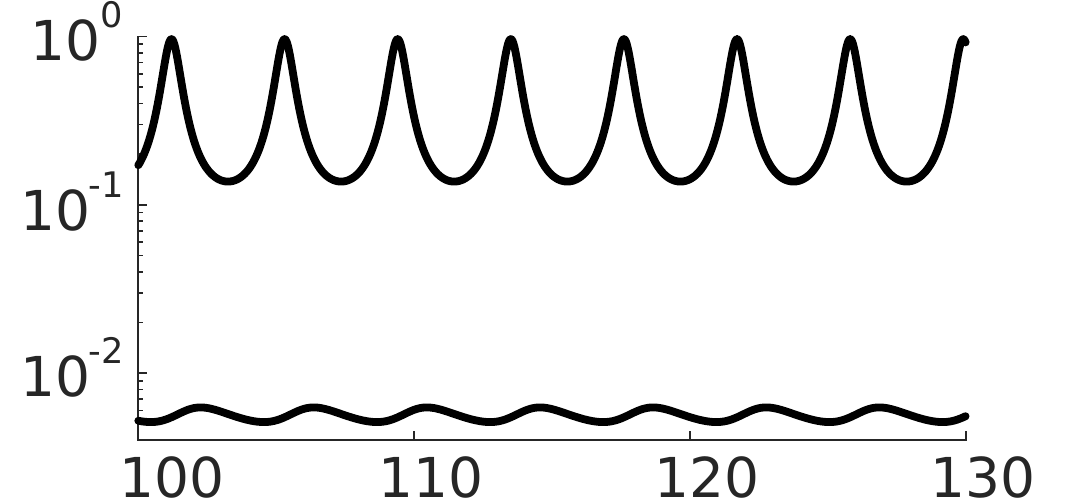}};
   \node[rotate=90, color=black] at (-2, 0) {$r_{1,2}$};
   \node[rotate=0] at (0, -1.25) {$t$};
   \node[rotate=0] at (0, -0.2) {\stateQSo\, / \stateSQo};
   \node[rotate=0] at (-1.1, 0.9) {$r_{1,2}$};
   \node[rotate=0] at (-1.1, -0.45) {$r_{2,1}$};
    \node[rotate=0] at (-2, 1.) {d)};
    \end{tikzpicture}
   \caption{
   \label{fig:observedStates}
   Asymptotic dynamic behaviors for the two population model after transients. Low and high levels of the firing rate $r(t)$ indicate quiescent (\stateQ) or spiking (\stateS) behavior.
   a) Symmetric (stable) equilibrium where both populations are quiescent (\stateQQ).
   b) Symmetric (stable)  equilibrium where both populations are spiking (\stateSS) (Transients leading up to the equilibrium are oscillatory).
   c) Asymmetric (stable) equilibrium where one population is quiescent and the other spiking (\stateQS\, or \stateSQ) (Transients leading up to the equilibrium are oscillatory only for the spiking population).
   d) Asymmetric (stable) limit cycle (\stateQSo\, or \stateSQo) where both populations oscillate with the same period (but with different amplitudes).
   The coupling parameter is $\kappa=1.8$ for panels a) through c) and $\kappa = 2.2$ for panel d); parameters $a=0.25, \hat{\eta}=-1, \Delta = 0.01, s=1$ are fixed throughout.
}
 \end{figure}
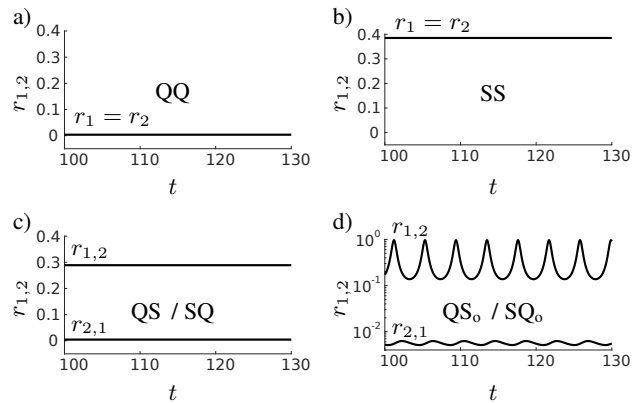

It is instructive to begin the analysis by surveying the possible asymptotic dynamic behavior for the firing rates $r_1$ and $r_2$ (or equivalently, $Z_1$ and $Z_2$) in the FRE~\eqref{eq:FREM1} and~\eqref{eq:FREM2} for $M=2$ populations. We may distinguish two types of asymptotic states as $t\rightarrow\infty$, namely  (i) \emph{symmetric states} characterized  by $r_1(t) = r_2(t)$ and $v_1(t)=v_2(t)$;  and
(ii) \emph{asymmetric states} characterized by $r_1(t) \neq r_2(t)$ and $v_1(t)\neq v_2(t)$. Furthermore, each neuron population may be in a state of \emph{quiescence} (\stateQ) or \emph{spiking} (\stateS), depending on whether $r_\sigma$ reflects low or high firing activity, respectively. In an asymmetric limit cycle, both populations oscillate around a distinct value corresponding to quiescence or spiking, respectively. Fig.~\ref{fig:observedStates}\ illustrates the possible asymptotic states that may be observed, depending on parameter values and initial conditions chosen.

 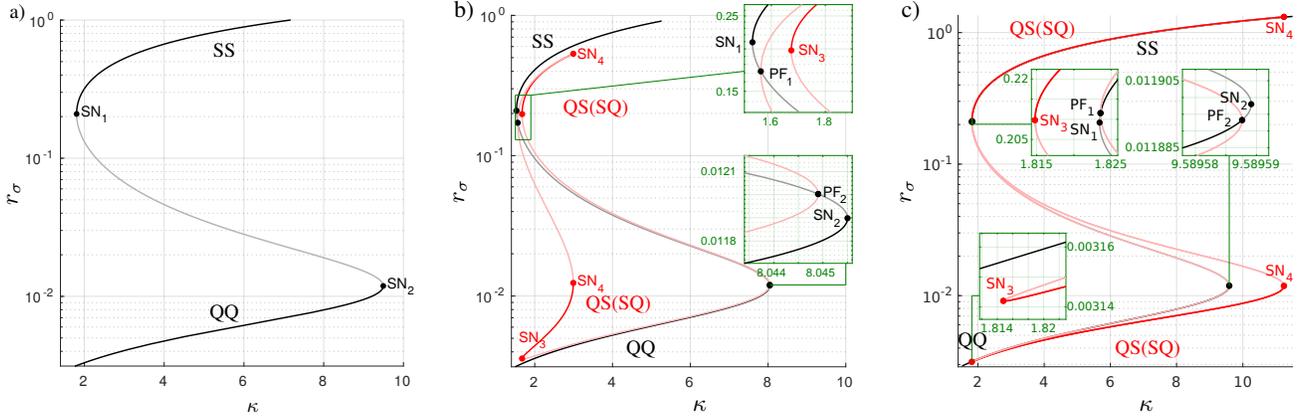
\begin{figure*}[htp!]
   \centering
    \begin{tikzpicture}
    \draw (0, 0) node[inner sep=0] {\includegraphics[width=0.32\textwidth]{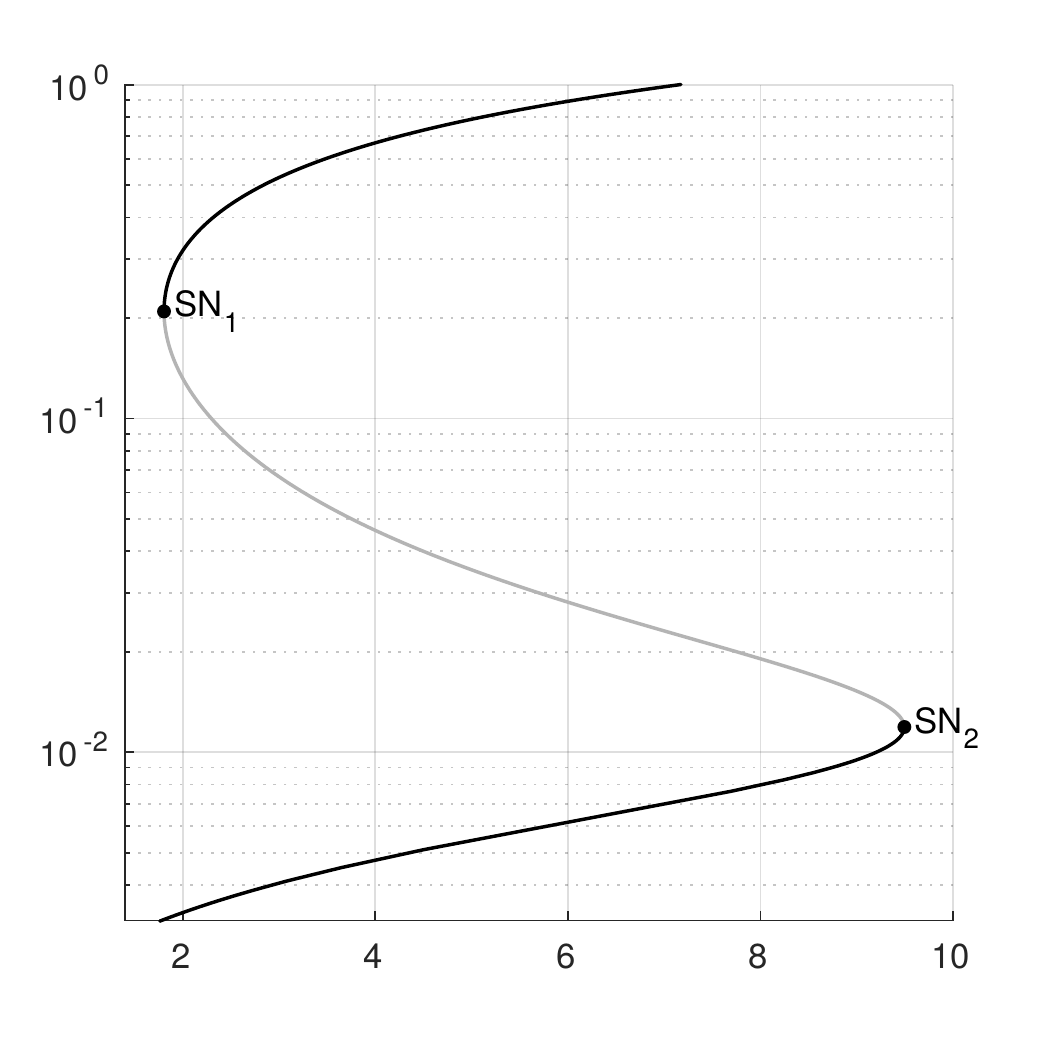}};
    \node[rotate=0,scale=0.85] at (0, 2.) {\stateSS};
    \node[rotate=0,scale=0.85] at (0, -1.5) {\stateQQ};
    \node[rotate=90, color=black] at (-2.8, .1) {$r_{\sigma}$};
    \node[rotate=0,scale=0.85] at (0.38, -2.75) {$\kappa$};
    \node[rotate=0,scale=0.85] at (-2.75, 2.5) {a)};
    \end{tikzpicture}
    \begin{tikzpicture}
    \draw (0, 0) node[inner sep=0] {\includegraphics[width=0.32\textwidth]{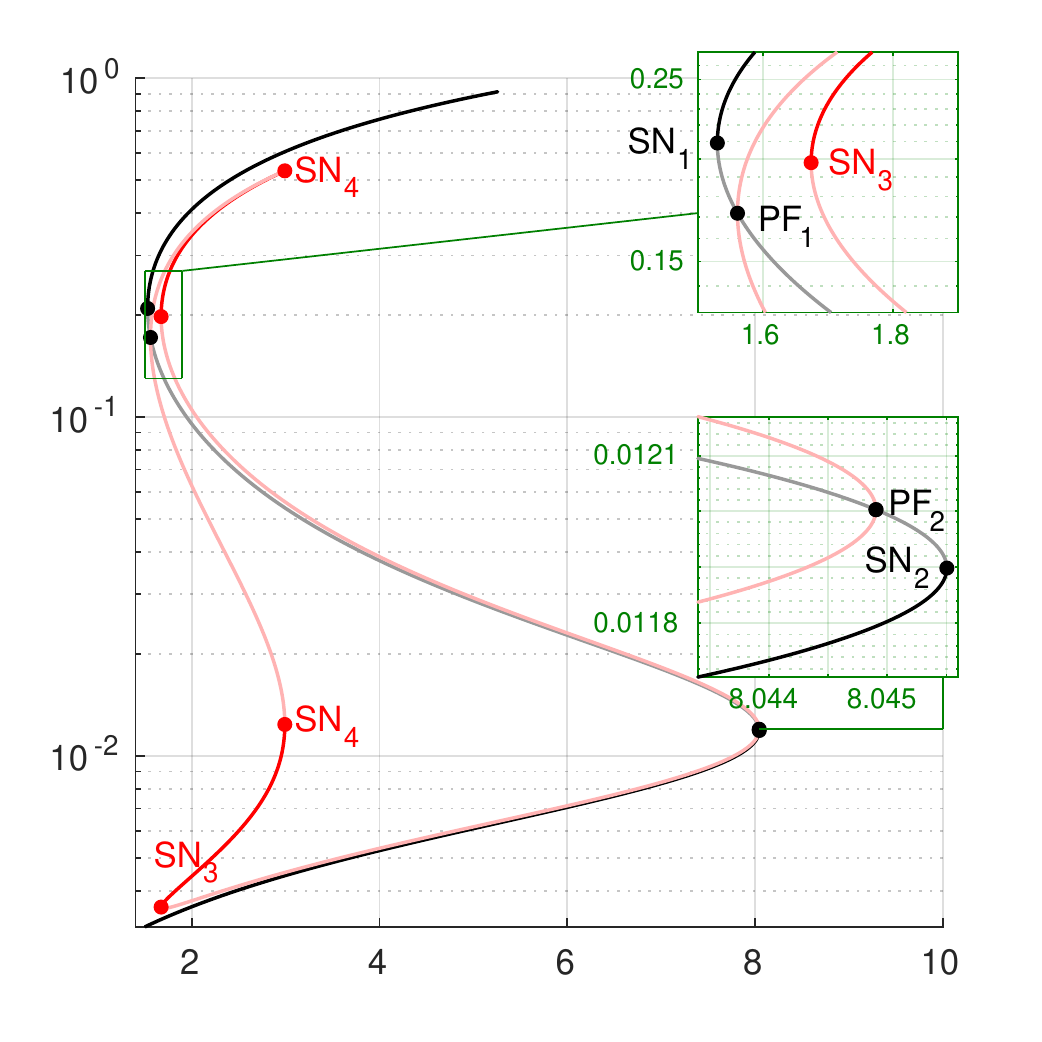}};
    \node[rotate=0,scale=0.85] at (-1.7, 2.1) {\stateSS};
    \node[rotate=0,scale=0.85] at (-0.4, -2) {\stateQQ};
    \node[rotate=0,color=red,scale=0.85] at (-1, 1.2) {\stateQS (\stateSQ)};
    \node[rotate=0,color=red,scale=0.85] at (-0.7, -1.4) {\stateQS (\stateSQ)};
    \node[rotate=90, color=black] at (-2.8, .1) {$r_{\sigma}$};
    \node[rotate=0] at (0.38, -2.75) {$\kappa$};
    \node[rotate=0] at (-2.75, 2.5) {b)};
    \end{tikzpicture}
    \begin{tikzpicture}
    \draw (0, 0) node[inner sep=0] {\includegraphics[width=0.32\textwidth]{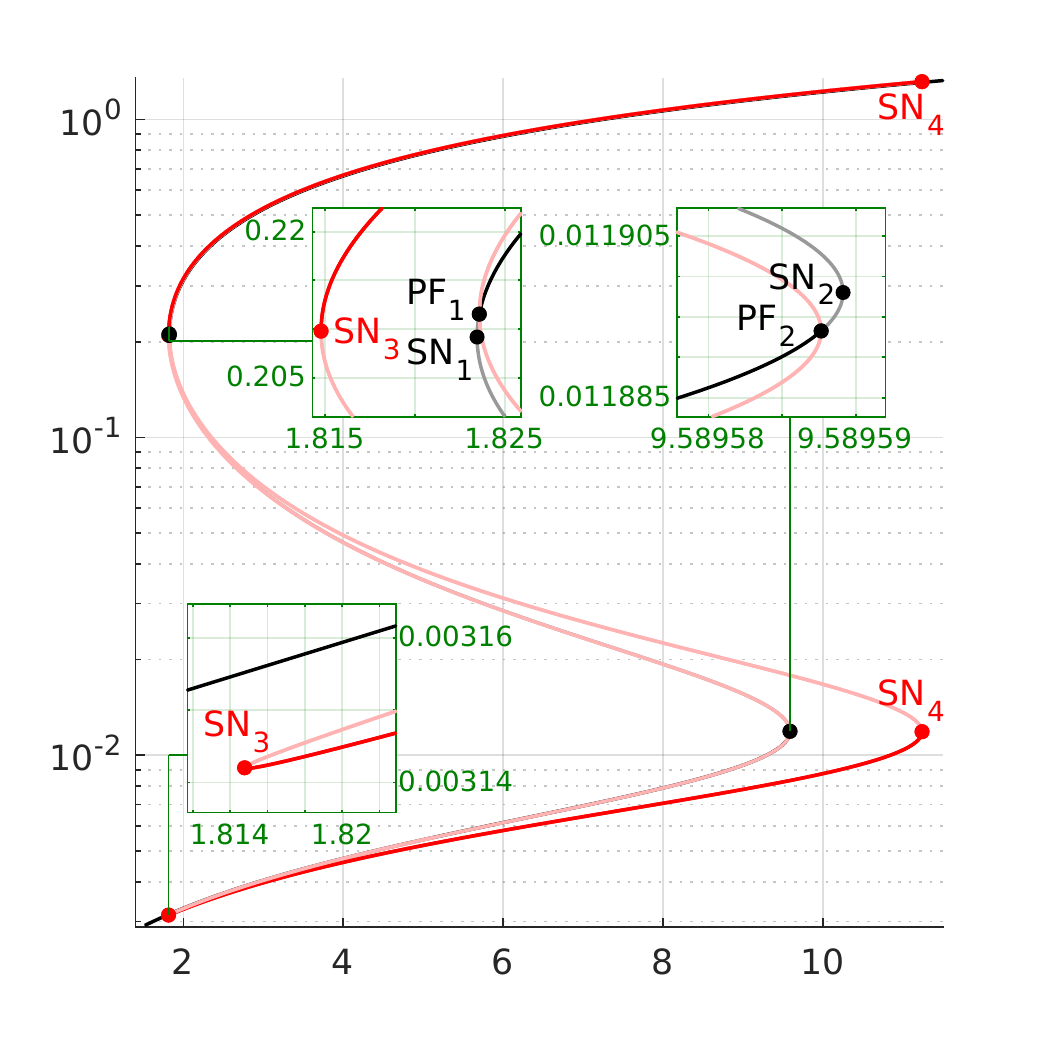}};
    \node[rotate=0,scale=0.85] at (.4, 2.) {\stateSS};
    \node[rotate=0,scale=0.85] at (-1.93, -1.9) {\stateQQ};
    \node[rotate=0,color=red,scale=0.85] at (-1, 2.25) {\stateQS (\stateSQ)};
    \node[rotate=0,color=red,scale=0.85] at (0.4, -2.) {\stateQS (\stateSQ)};
    \node[rotate=90, color=black] at (-2.8, .1) {$r_{\sigma}$};
    \node[rotate=0] at (0.38, -2.75) {$\kappa$};
    \node[rotate=0] at (-2.75, 2.5) {c)};
    \end{tikzpicture}
   \caption{
   \label{fig:bifurcationdiagramOne}
   Bifurcation diagrams for $M=2$ populations in $\kappa$ for the firing rate $r_\sigma$ reveal symmetric equilibria, $r_1=r_2$ (black, gray) and asymmetric equilibria, $r_1\neq r_2$, (red, light red), emerging in bifurcations as follows.
   a) $a=0$: Both populations are decoupled. Symmetric equilibria are folded in two saddle-node bifurcations (\SNone,  \SNtwo ) where the lower and upper branches (black) corresponding to (Q)uiescence and (S)piking, respectively, are stable; the middle branch (gray) is unstable. Thus, four states are possible, (\stateSS, \stateQQ, \stateQS\ and \stateSQ), facilitating  multistability and hysteretic behavior.
   b) $a=0.18$: Symmetric equilibria with $r_1=r_2$ seen for $a=0$ are still present (black, gray). However, the unstable branch (gray) undergoes a pitchfork bifurcation in \PFone\ and \PFtwo, giving rise to unstable asymmetric equilibria \stateQS\ / \stateSQ\ with $ r_1 \neq r_2 $ (light red).
   These equilibria are connected in a loop, folded twice in two saddle-node bifurcations (\SNthree\ and \SNfour), giving rise to the co-existence of two stable asymmetric states (red) where one population is quiescent while the other is spiking, (\stateSQ\ or \stateQS). c) $a=-0.01$: For $a<0$ the order of \SNthree,  \SNfour\ and \SNone, \SNtwo, \PFone, \PFtwo\ is reversed along the $\kappa$ direction when compared to $a>0$.
   Parameters are $\Delta=0.01, \hat{\eta}=-1, s=1$ everywhere. Chosen values for $a$ are indicated as dashed horizontal lines in Fig.~\ref{fig:stabilitydiagram}.
   }
 \end{figure*}

Solution branches reported in Fig.~\ref{fig:bifdiag_M1} for $M=1$ population translate to symmetric states in the model with $M=2$ populations.
To see this, let us first consider two special parameter choices: $a=0$ (decoupled populations) and $a=1$ (two populations effectively act like one large population). In these cases, the system with $M=2$ populations displays the same bifurcation behavior as $M=1$ population, as shown in Fig.~\ref{fig:bifurcationdiagramOne}a) for $a=0$.
The branch with low firing rate (\stateQQ) corresponds to quiescent neurons with coherent stationary phases; whereas the branch with high firing rate (\stateSS) corresponds to spiking populations whose synchronization level and firing rate grow with increasing coupling strength $\kappa$. Just as for $M=1$ population, the system exhibits bistable regions in which both configurations, (Q)uiescence and (S)piking, are possible.
However, note that in the case of $a=1$, both populations may \textit{only} attain identical (symmetric) configurations of quiescence or spiking, namely \stateSS\ or \stateQQ;  in contrast, the decoupled case with $a=0$  additionally and trivially allows for the two populations to attain distinct (asymmetric) configurations, namely, \stateSQ\ or \stateQS.
Importantly, symmetric states persist even when $a\neq0$ or $a\neq1$ since parameters are symmetric across the two populations.
Specifically, if $r$ is an equilibrium of the $ M=1 $ population system, then so is $(r,r)$  an equilibrium of the $M=2$ population system but now with $ \kappa $ replaced by $\kappa/(1+a)$. For this reason, the solution branches of symmetric equilibria seen for $M=1$ translate to the $M=2$ system, including the saddle-node bifurcations \SNone\ and \SNtwo\ seen in Fig.~\ref{fig:bifdiag_M1}, in between which, for $a\geq0$, two stable symmetric states \stateSS\ and \stateQQ\ co-exist. However, for $a<0$, the region of bistability for symmetric states is bounded by the pitchfork bifurcations \PFone\ and \PFtwo\ rather than \SNone\ and \SNtwo, since both branches emanating from the  bifurcation point \SNone\ (\SNtwo) are unstable, one of them gaining stability at \PFone\ (\PFtwo), see Figs.~\ref{fig:bifurcationdiagramOne}c) and~\ref{fig:stabilitydiagram}a). \PFone\ and \PFtwo\ also give rise to asymmetric states, as we explain in the following.

Asymmetric states, corresponding to  \stateQS\ or \stateSQ\ configurations with $r_1\neq r_2$, are trivially possible when the two populations are decoupled ($a=0$); however, their range of existence and stability off the degenerate cases $a\neq0$ and $a\neq 1$ deserves further exploration, and we consider small perturbations for $a\neq0$.
Considering the case of $a=0.18$ in Fig.~\ref{fig:bifurcationdiagramOne}b) we observe that unstable asymmetric states (light red) branch off the unstable symmetric state (gray) in pitchfork bifurcations \PFone\ and  \PFtwo\, (see also inset). The set of asymmetric equilibria forms a loop in $(\kappa, r)$-space with two folds, i.e., the equilibria undergo saddle-node bifurcations in \SNthree\ and \SNfour, between which asymmetric states are stable. As a result, the \stateQS\ and \stateSQ\ (red) emerge as bistable  asymmetric configurations. These stable asymmetric branches may co-exist with the bistable symmetric solution branches \stateQQ\ and \stateSS\, (black).
Note that for $a>0$, \PFone\ and \PFtwo\ lie outside \SNthree\ and \SNfour, while for $a<0$ \PFone\ and \PFtwo\ lie inside \SNthree\ and \SNfour; as a consequence, the
existence of asymmetric states is bounded by \PFone\ and \PFtwo\ for $a>0$ and \SNthree\ and \SNfour\ for $a<0$, see Fig.~\ref{fig:bifurcationdiagramOne}b) and c).
Moreover, for $a > 0$, asymmetric states exist only in a relatively narrow range of intermediate coupling strength $\kappa$; by contrast, for $a \leq 0$, the $\kappa-$range of existence of asymmetric equilibria rapidly expands as the value of $a$ decreases (see Fig.~\ref{fig:stabilitydiagram}a).

Note that, considering the case of decoupled populations with $a=0$, symmetric and asymmetric branches may appear like they coincide when inspecting Fig.~\ref{fig:bifurcationdiagramOne}, however, the two types of solution branches are not identical: While the projections $Z_1\in \C$ and $Z_2\in\C$ indeed share identical values for symmetric and asymmetric equilibria, this cannot hold true in the full phase space for  $(Z_1,Z_2)\in\C^2$, where the definitions for symmetric ($r_1 = r_2, v_1=v_2$) and asymmetric states ($r_1\neq r_2, v_1\neq v_2$) are obeyed.

\subsection{Birth and destruction of limit cycle oscillations}

\begin{figure*}[htp!]
\centering
\begin{tikzpicture}
    \draw(0,0) node[inner sep=0]{\includegraphics[width=0.45\textwidth]{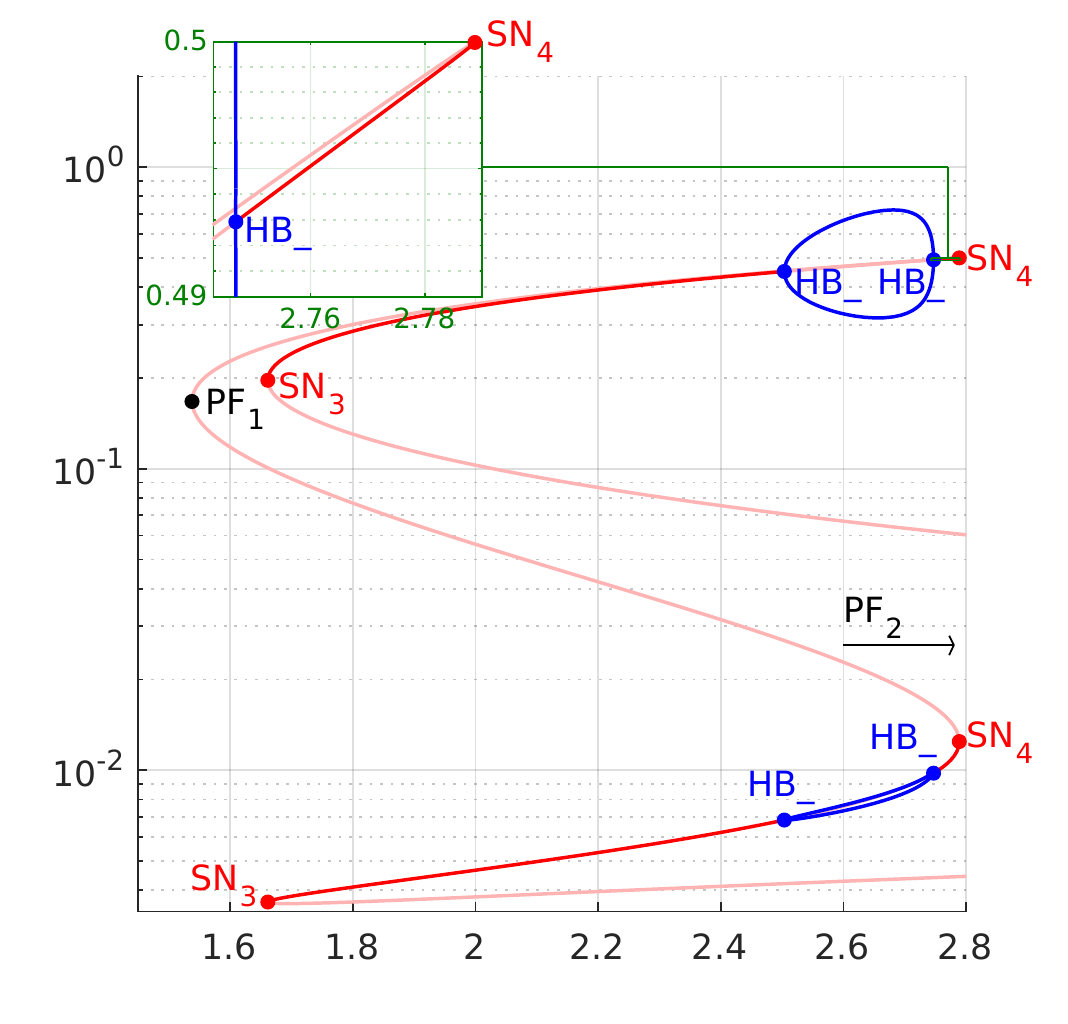}};
    \node[rotate=90, color=black] at (-4, .2) {$\text{min}_t (r_{\sigma}), \text{max}_t (r_{\sigma})$};
    \node[rotate=0] at (0, -3.75) {$\kappa$};
    \node[rotate=0] at (-4, 3.5) {a)};
   \end{tikzpicture}
   \begin{tikzpicture}
    \draw (0, 0) node[inner sep=0] {\includegraphics[width=0.45\textwidth]{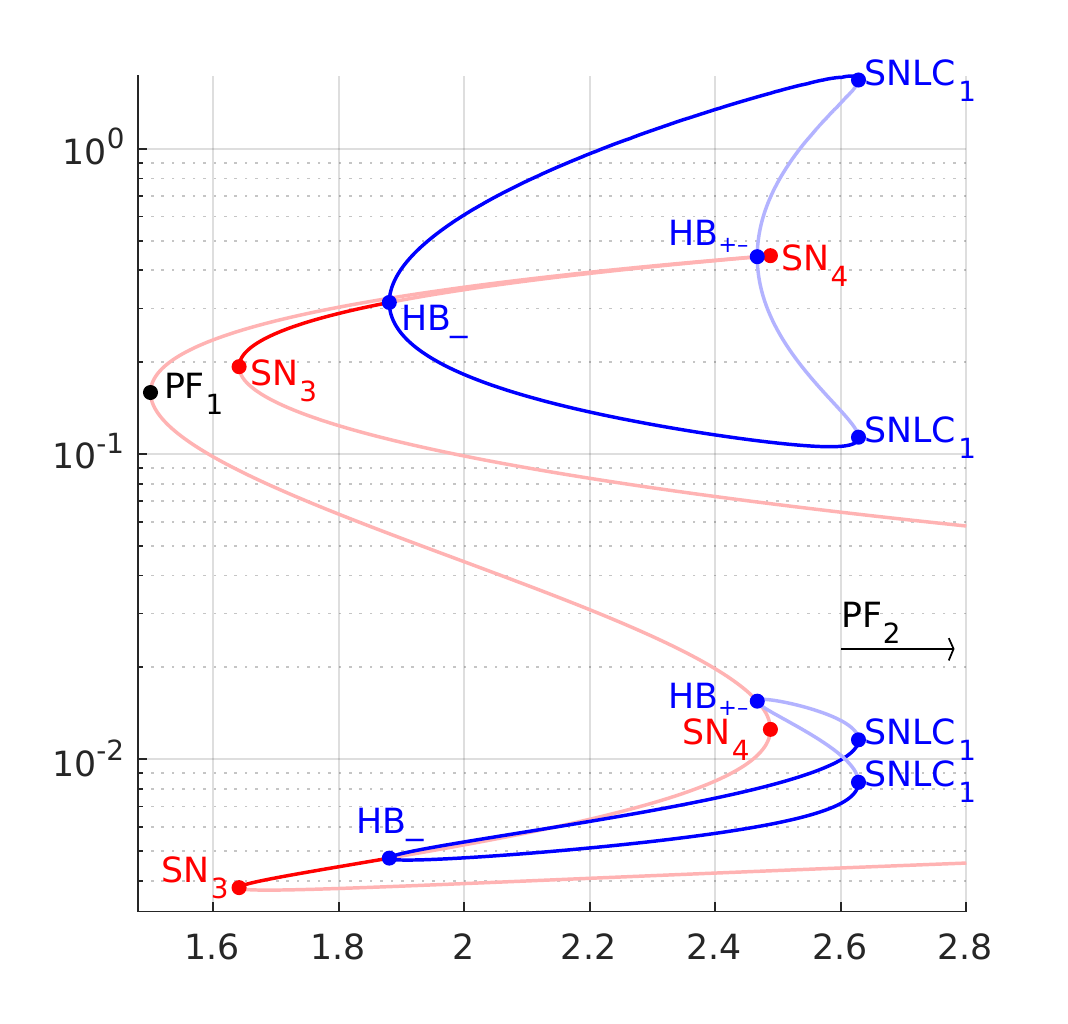}};
    \node[rotate=90, color=black] at (-4, .2) {$\text{min}_t (r_{\sigma}), \text{max}_t (r_{\sigma})$};
    \node[rotate=0] at (0, -3.75) {$\kappa$};
    \node[rotate=0] at (-4, 3.5) {b)};
   \end{tikzpicture}
    \begin{tikzpicture}
    \draw (0, 0) node[inner sep=0] {\includegraphics[width=0.45\textwidth]{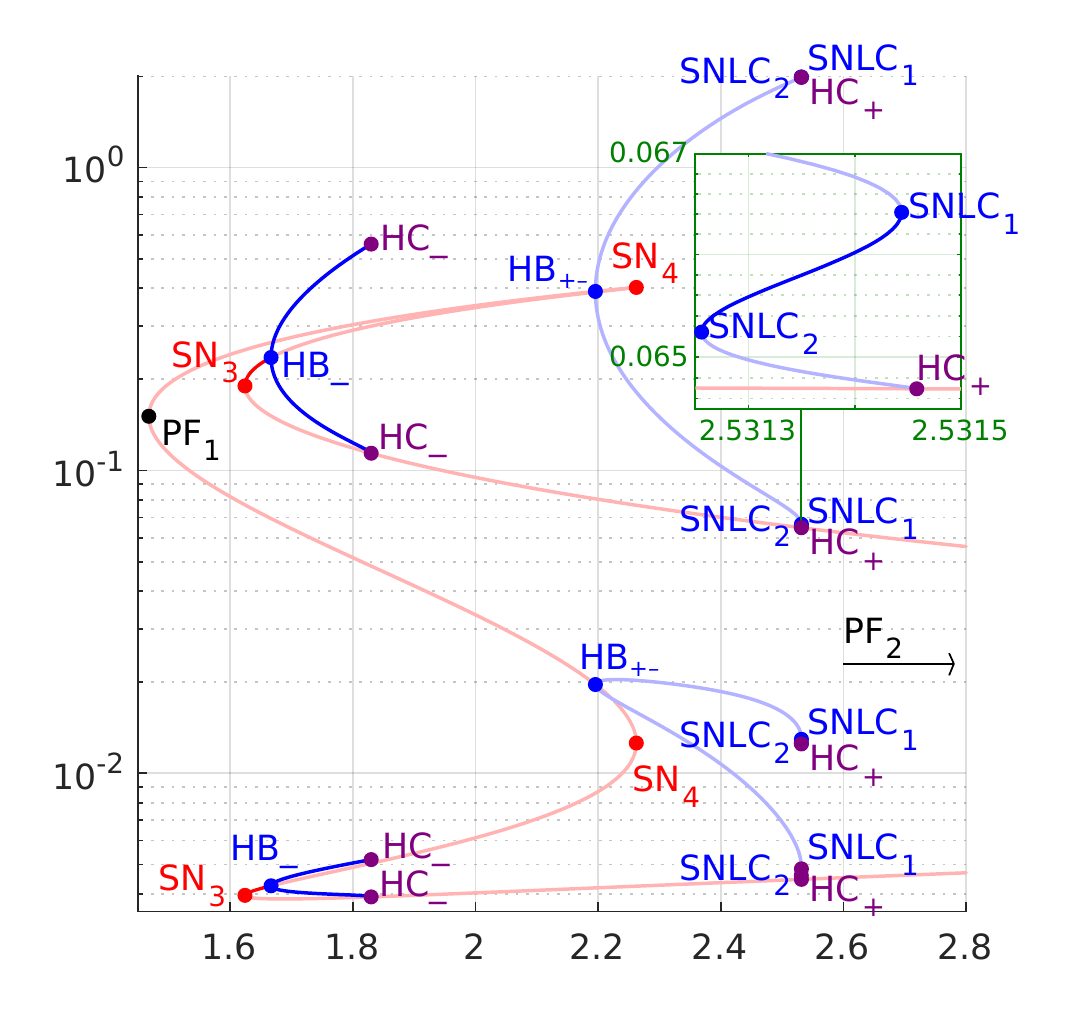}};
    \node[rotate=90, color=black] at (-4, .2) {$\text{min}_t (r_{\sigma}), \text{max}_t (r_{\sigma})$};
    \node[rotate=0] at (0, -3.75) {$\kappa$};
    \node[rotate=0] at (-4, 3.5) {c)};
    \end{tikzpicture}
    \begin{tikzpicture}
    \draw(0,0) node[inner sep=0]{\includegraphics[width=0.45\textwidth]{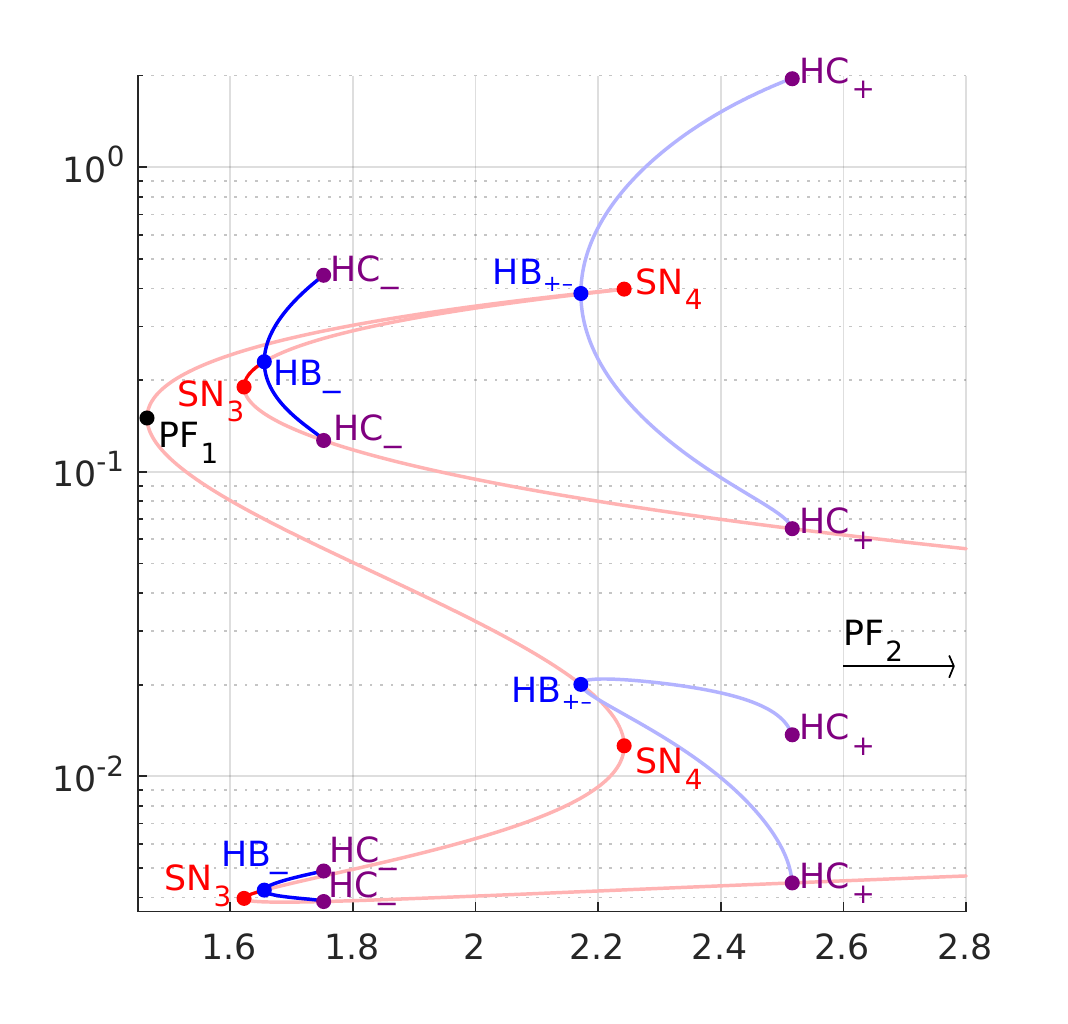}};
    \node[rotate=90, color=black] at (-4, .2) {$\text{min}_t (r_{\sigma}), \text{max}_t (r_{\sigma})$};
    \node[rotate=0] at (0, -3.75) {$\kappa$};
    \node[rotate=0] at (-4, 3.5) {d)};
    \end{tikzpicture}
   \caption{
   \label{fig:bifurcationdiagramTwo}
   Birth and destruction of asymmetric limit cycle oscillations (blue) in the firing rate for $M=2$ populations while varying $\kappa$.
   Possible bifurcation scenarios are illustrated for various values of $a$ (also indicated as dashed lines in Fig.~\ref{fig:stabilitydiagram}).
   a) $a=0.204$: stable limit cycles are born in the supercritical Hopf bifurcation \HBstab (the bifurcation curve is intersected twice, see Fig.~\ref{fig:stabilitydiagram}).
   b) $a=0.25$: stable and unstable limit cycles are born in \HBstab\ and \HBustbstab, respectively, and annihilate in a saddle-node of limit cycles bifurcation (\SNLCone).
   c) $a=0.297$: stable and unstable limit cycles are destroyed in a homoclinic bifurcation \HCstab\ and \HCustb, respectively. Moreover, the unstable cycle is subject to two saddle-node-of-limit-cycles bifurcations at \SNLCone\ and \SNLCtwo, in between which it gains stability (see inset); the unstable limit cycle branch emerging from \SNLCtwo\ gets destroyed in \HCustb.
   d) $a=0.302$: the unstable limit cycle born in \HBustbstab\ is destroyed in the homoclinic \HCustb\ ; saddle-node-of-limit-cycles bifurcations are now absent.
   Symmetric equilibria connecting to \PFone\, and \PFtwo\, are omitted for simplicity.
   Stable and unstable solution branches are shown as dark and light colored shades, respectively.
   Parameters are $\Delta=0.01, \hat{\eta}=-1, s=1$ everywhere.
   }
 \end{figure*}
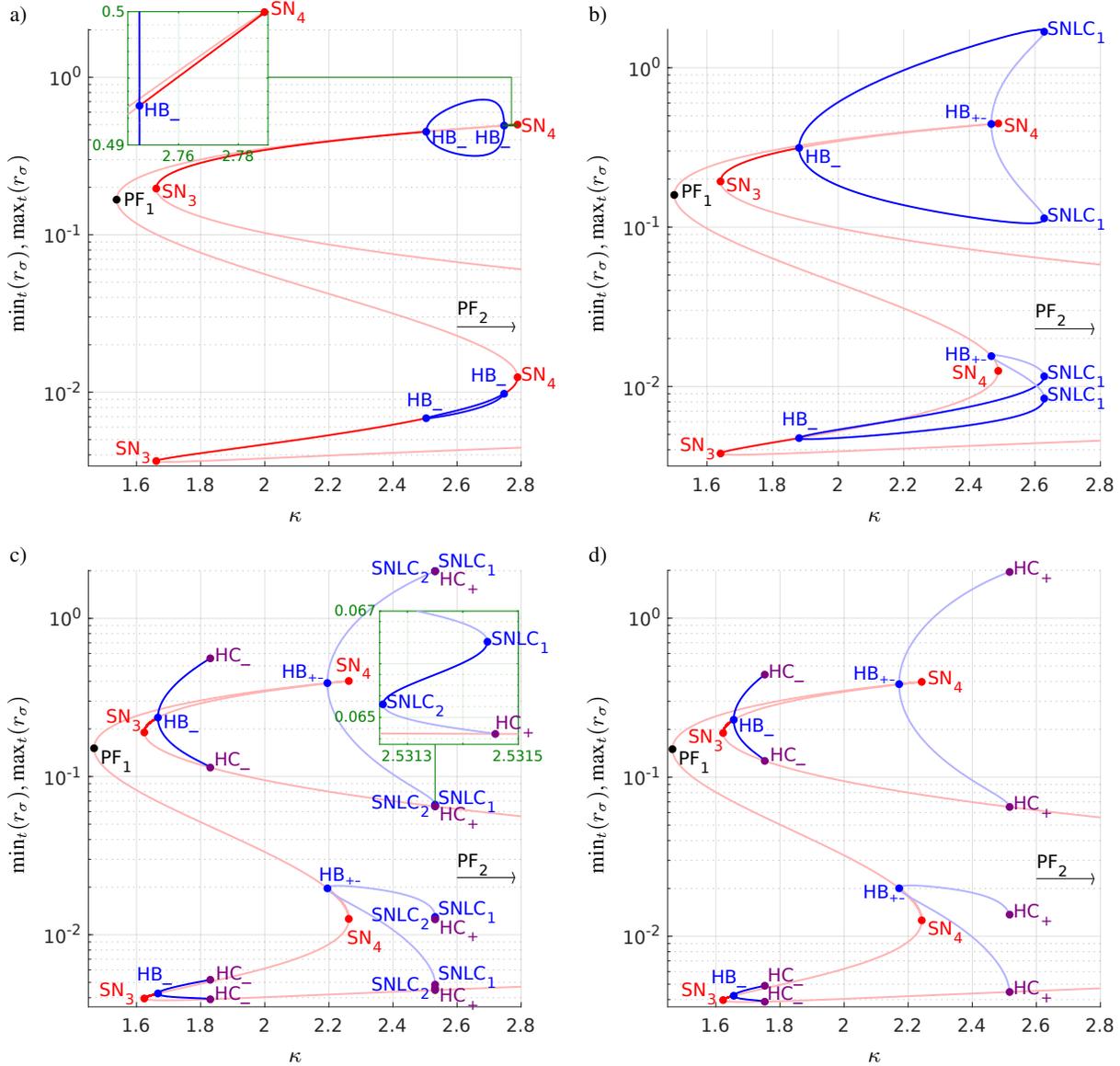

For larger values of the inter-coupling strength, $a$, asymmetric equilibria  \stateQS\ (\stateSQ) may undergo Hopf bifurcations giving rise to limit cycle oscillations (\stateQSo, \stateSQo), indicated by their minima/maxima (blue) in Fig.~\ref{fig:bifurcationdiagramTwo}a) through d). Since these limit cycles branch off asymmetric equilibria (red), they correspond to asymmetric configurations characterized by firing rates $r_1(t)\neq r_2(t)$. These limit cycles are created and destroyed in various bifurcations, as outlined in the following.

\paragraph{Birth of stable limit cycles (\HBstab).}
Stable limit cycles (blue minima/maxima) are born in the supercritical Hopf bifurcation denoted by \HBstab, as shown in Fig.~\ref{fig:bifurcationdiagramTwo}a) for $a=0.204$. As $\kappa$ increases, the amplitude waxes and wanes, as the bifurcation \HBstab\ is intersected twice in the direction of varying $\kappa$, see also Fig.~\ref{fig:stabilitydiagram}).

\paragraph{Birth of stable/unstable limit cycles and annihilation in saddle-node-of-limit-cycle bifurcation (\HBstab,\HBustbstab,\SNLCone).}
Stable limit cycles (blue) are still born in a supercritical Hopf bifurcation at \HBstab, but now an unstable limit cycle (light blue) of smaller amplitude emerges for greater $\kappa$ in the supercritical Hopf (with repelling center manifold) at \HBustbstab.
The continuum of cycles folds over in a \emph{saddle-node of limit cycles} bifurcation at \SNLCone, where the stable and unstable limit cycles coalesce and disappear, see Fig.~\ref{fig:bifurcationdiagramTwo}b) for $a=0.25$.

\paragraph{Stabilization of unstable limit cycle in secondary saddle-node-of-limit-cycle bifurcation (\SNLCtwo).}\label{ref:stabilization_unstable_limit_cycle}
Stable and unstable limit cycles are created in \HBstab\ and \HBustbstab. While the stable limit cycle is destroyed in the homoclinic bifurcation \HCstab, the unstable limit cycle is subject to a more complicated series of bifurcations:  It undergoes not only one, but two saddle-node of cycles bifurcations, \SNLCtwo\ and  \SNLCone. The unstable limit cycle emerging from \SNLCtwo\ collides with the saddle equilibrium of the asymmetric branch in the \emph{homoclinic} bifurcation \HCustb\ and is destroyed,  as shown in Fig.~\ref{fig:bifurcationdiagramTwo}c) for $a=0.297$.

\paragraph{Simple birth and destruction of stable/unstable limit cycles (\HBstab, \HCstab, \HBustbstab, \HCustb).}
The stable and unstable limit cycles are born in the Hopf bifurcations \HBstab\ and \HBustbstab\ and are destroyed in the homoclinic bifurcations \HCstab\ and  \HCustb, respectively. The complicated scenario including two saddle-node-of-limit-cycles bifurcations from \ref{ref:stabilization_unstable_limit_cycle} is entirely absent. This simple scenario is shown in Fig.~\ref{fig:bifurcationdiagramTwo}d) for $a=0.302$.

\subsection{Stability diagram}

We now explain how the various bifurcation scenarios are related, i.e., how stability boundaries  are connected in the $(\kappa,a)$-parameter plane and how bifurcation curves are structured around bifurcation points of higher co-dimension.

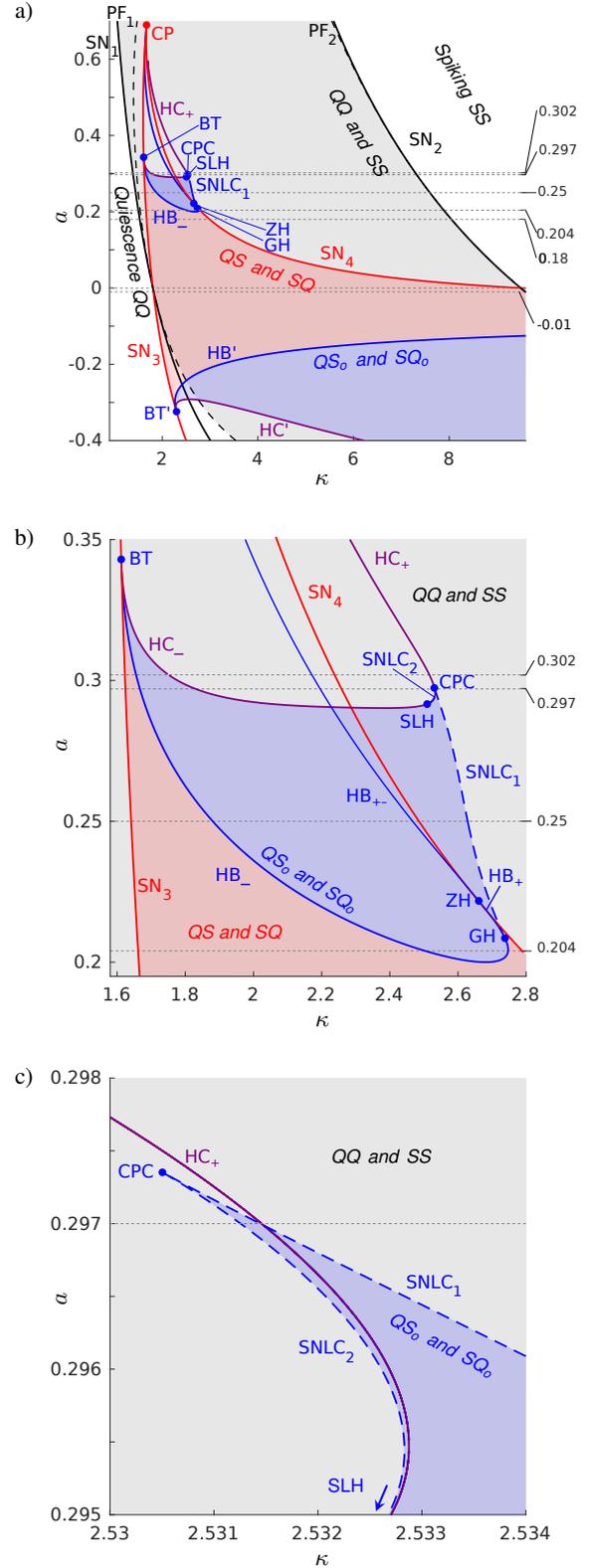
\begin{figure}[htp!]
 \centering
   \begin{tikzpicture}
    \draw(0,0) node[inner sep=0]{\includegraphics[width=0.4\textwidth]{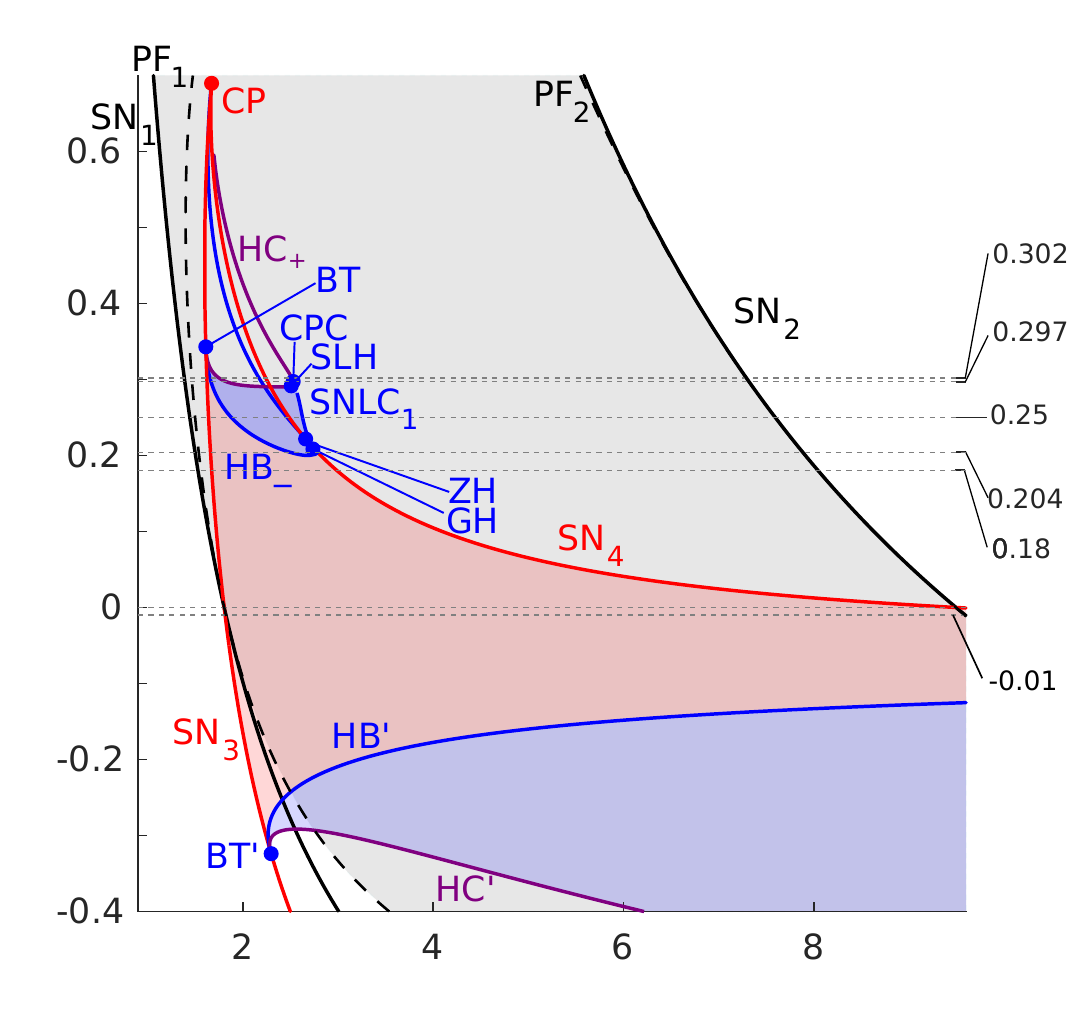}};
    \node[rotate=-60,text width = 6em] at (2.2, 1.8) {{\scriptsize \it \fontfamily{phv}\selectfont Spiking \stateSS}};
    \node[rotate=-60,text width = 6em] at (0.8, 1.2) {{\scriptsize \it \fontfamily{phv}\selectfont \stateQQ\ and  \stateSS}};
    \node[rotate=-20,color=red,text width = 6em] at (-0.3, -0.5) {{\scriptsize \it \fontfamily{phv}\selectfont \stateQS\ and \stateSQ}};
    \node[rotate=3,color=blue,text width = 6em] at (1, -1.6) {{\scriptsize \it \fontfamily{phv}\selectfont \stateQSo\ and \stateSQo}};
    \node[rotate=280,text width = 6em] at (-2.35, -0.2) {{\scriptsize \it \fontfamily{phv}\selectfont Quiescence \stateQQ}};
    \node[rotate=90, color=black] at (-3.3, 0.3) {$a$};
    \node[rotate=0] at (0.18, -3.2) {$\kappa$};
    \node[rotate=0] at (-3.8, 3.05) {a)};
   \end{tikzpicture}
   \begin{tikzpicture}
    \draw(0,0) node[inner sep=0]{\includegraphics[width=0.4\textwidth]{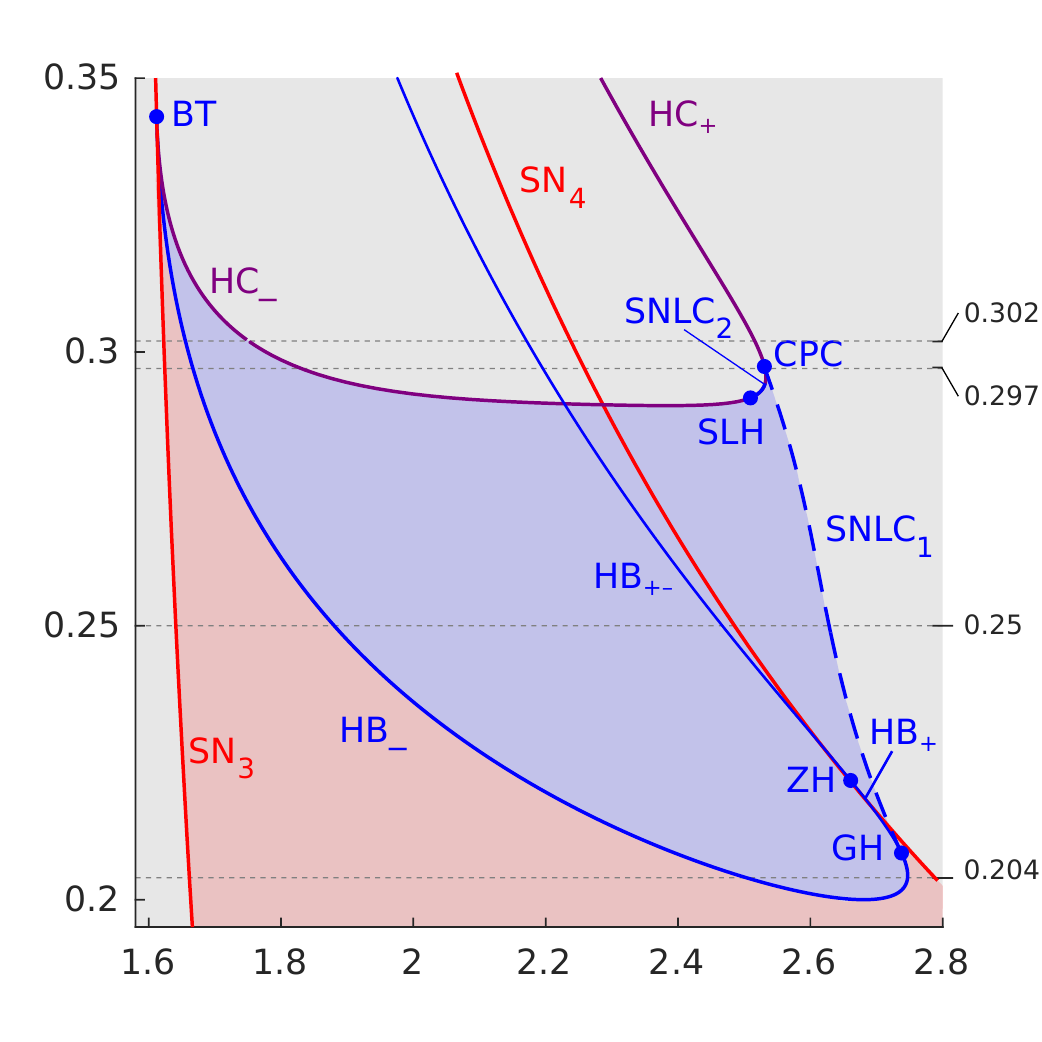}};
    \node[rotate=0] at (2, 2.3) {{\scriptsize \it \fontfamily{phv}\selectfont \stateQQ\ and \stateSS}};
    \node[rotate=-32,color=blue] at (0, -1.5) {{\scriptsize \it \fontfamily{phv}\selectfont \stateQSo\ and \stateSQo}};
    \node[rotate=0,color=red] at (-1, -2.2) {{\scriptsize \it \fontfamily{phv}\selectfont \stateQS\ and \stateSQ}};
    \node[rotate=90, color=black] at (-3.3, 0.3) {$a$};
    \node[rotate=0] at (0.18, -3.4) {$\kappa$};
    \node[rotate=0] at (-3.8, 3.05) {b)};
   \end{tikzpicture}
   \begin{tikzpicture}
    \draw(0,0) node[inner sep=0]{\includegraphics[width=0.4\textwidth]{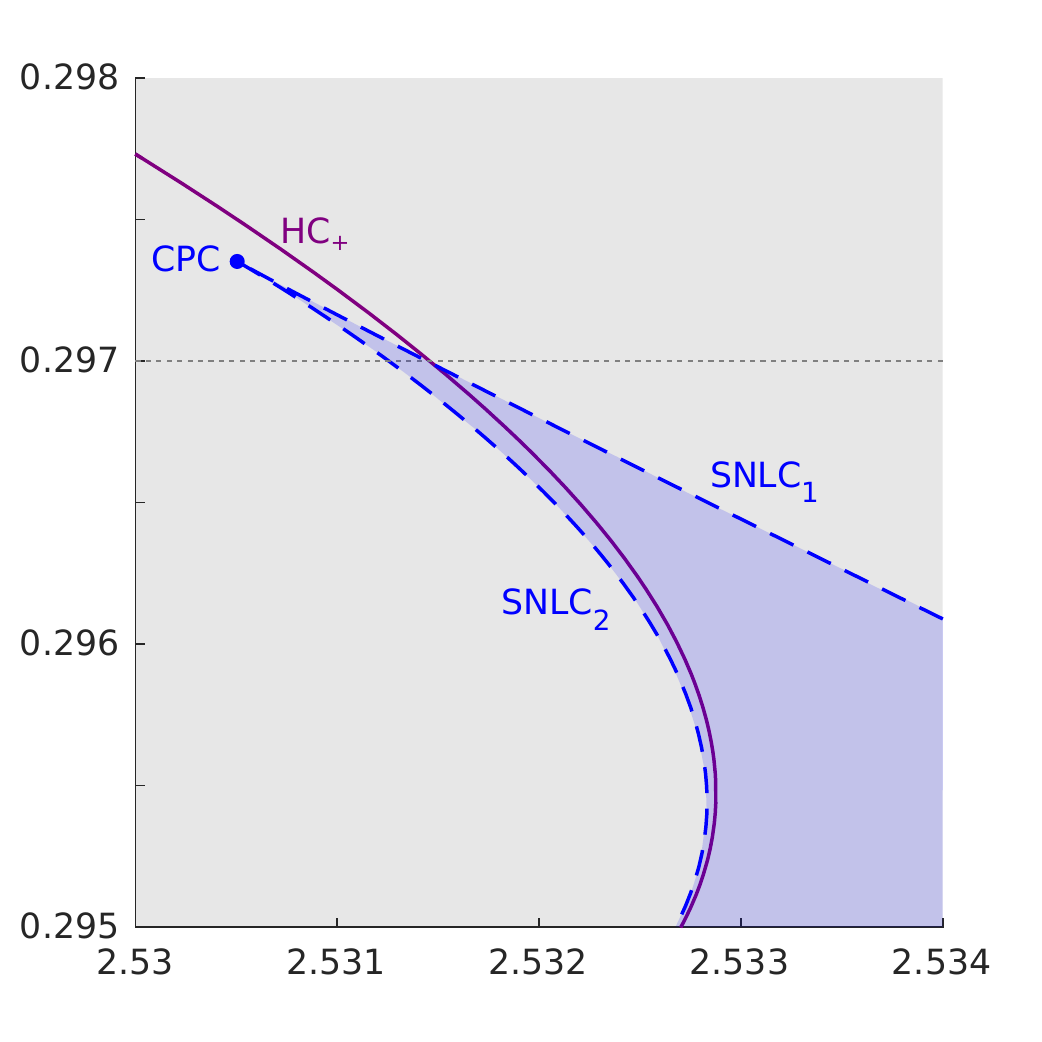}};
    \node[rotate=0,text width = 6em] at (1.25, 2.) {{\scriptsize \it \fontfamily{phv}\selectfont \stateQQ\ and \stateSS}};
    \node[rotate=-28,color=blue,text width=2.5cm] at (2.1, -0.7) {{ \scriptsize \it \fontfamily{phv}\selectfont \stateQSo\ and \stateSQo}};
    \node[rotate=0, color=blue] at (0.5,-2.4) {\scriptsize \SLH};
    \draw[arrow,color=blue] (1.05,-2.4) -- (0.9,-2.75);
    \node[rotate=90, color=black] at (-3.3, 0.1) {$a$};
    \node[rotate=0] at (0.18, -3.4) {$\kappa$};
    \node[rotate=0] at (-3.8, 3.05) {c)};
   \end{tikzpicture}
\caption{
\label{fig:stabilitydiagram}
Stability diagram for $M=2$ populations in the $(\kappa,a)$-parameter plane with $\Delta=0.01, \hat{\eta}=-1, s=1$ fixed.
Gray shading indicates a bistable region of the two stable symmetric equilibria, \stateQQ\ and \stateSS.
Red shading indicates the bistable region of stable asymmetric equilibria, that is, \stateQS, or, equivalently, \stateSQ.
Blue shading indicates the bistable region of stable asymmetric limit cycles (\stateQSo\ and \stateSQo).
Dashed black lines delineate the choices of parameter $a$ for the bifurcation diagrams in Fig.~\ref{fig:bifurcationdiagramOne} and Fig.~\ref{fig:bifurcationdiagramTwo}.
}
\end{figure}

Let us first consider the overall bifurcation structure for a larger parameter range ($\kappa,a$) as displayed in Fig.~\ref{fig:stabilitydiagram}a), mainly focusing on symmetric (\stateQQ, \stateSS) and asymmetric equilibria \stateQS\ (or \stateSQ). On the branches of \emph{symmetric} equilibria (\stateQQ, \stateSS) two saddle-node bifurcations occur, \SNone\, and \SNtwo\ (black), which coalesce in a codimension 2 cusp point for large $a$ (not shown). The gray shaded region of bistability between \stateQQ\ and \stateSS\ is bounded by \SNone\ and \SNtwo\ for $a\geq0$ and by \PFone\ and \PFtwo\ (dashed black curves in Fig.~\ref{fig:stabilitydiagram}a)) for $a<0$, respectively. Note that the curve \PFtwo\ lies very close to \SNtwo\ in the shown parameter range, $-0.4\leq a\leq0.7$.

Unstable saddle branches of the symmetric equilibria between (or outside) \SNone\ and \SNtwo\  undergo pitchfork bifurcations \PFone\ and \PFtwo, which give rise to unstable asymmetric branches (see light red curves in Fig.~\ref{fig:bifurcationdiagramOne}b) and c) and Fig.~\ref{fig:bifurcationdiagramTwo}a) through d)).
These unstable asymmetric branches gain stability on the saddle-node bifurcation curves \SNthree\ and \SNfour\ (red curves in Fig.~\ref{fig:stabilitydiagram}a)), which meet in the codimension 2 cusp point \CP. The resulting asymmetric stable configurations (\stateQS\ or \stateSQ) reside inside the red shaded region bounded by the saddle-node bifurcation curves \SNthree\ and \SNfour\ and the supercritical Hopf bifurcation curves \HBstab\ and \HBprime.

In \HBstab\ and \HBprime, stable asymmetric equilibria \stateQS\ and \stateSQ\ lose stability, resulting in stable asymmetric limit cycles \stateQSo\ (or \stateSQo) within the blue shaded regions;
these limit cycles may get destroyed in the homoclinic bifurcations denoted by \HCstab\ and \HCprime\ (violet). Hopf (\HBstab\ and \HBprime) and homoclinic bifurcation curves associated with the emergence and destruction of these limit cycles (\HCstab\ and \HCprime) meet with the (asymmetric) saddle-node bifurcation curve \SNthree\ in two other bifurcation points of codimension 2, namely the Bogdanov-Takens points \BT\ and \BTprime, respectively, characterized by double zero eigenvalues~\cite{Kuznetsov1998}.

The bifurcations pertaining to the asymmetric limit cycles are structured around further, more complicated bifurcation curves and bifurcation points of higher co-dimension, see Fig.~\ref{fig:stabilitydiagram}b) and c).
Following the Hopf bifurcation curve \HBstab\ in panel b), we arrive at a Generalized Hopf bifurcation point (\GH) of codimension 2~\cite{Kuznetsov1998,Guckenheimer2007GH}. Such a point not only has a pair of purely imaginary eigenvalues, but also the first Lyapunov coefficient for the Hopf bifurcation changes sign at this point so that subcritical (\HBustb) and supercritical (\HBstab) Hopf bifurcations are separated in \GH; in addition, a branch of  saddle-node of limit cycle bifurcations, \SNLCone, emerges from \GH\ where the stable and unstable limit cycles born in \HBstab\ and \HBustb\ are annihilated.

Following the bifurcation curve \HBustb,  the associated subcritical Hopf bifurcation tangentially intersects the saddle-node bifurcation \SNfour\ in the Zero-Hopf bifurcation \ZH\ (or saddle-node Hopf bifurcation)~\cite{Kuznetsov1998,Guckenheimer2007ZH}, characterized by a zero eigenvalue and a pair of purely imaginary eigenvalues.
At \ZH, the first Lyapunov coefficient vanishes once more and changes sign.
Hopf bifurcations \HBustbstab\ above the \ZH\ point are supercritical (i.e., having a negative first Lyapunov coefficient), but continue to produce unstable limit cycles as the center manifold (of the Hopf bifurcation) is repelling.

Following the saddle-node bifurcation of limit cycles curve \SNLCone, we observe that it terminates in another bifurcation point of codimension 2, Cusp of Cycles  (\CPC), where it collides with a second saddle-node bifurcation of limit cycles curve \SNLCtwo. This latter bifurcation curve merges with the homoclinic bifurcation curve \HCustb\ in a codimension $\geq2$ point \SLH, see Fig.~\ref{fig:stabilitydiagram}c).
The point \SLH\ separates two branches of the homoclinic curves, \HCstab\ and \HCustb, and tangentially intersects with \SNLCtwo. Homoclinic bifurcations on \HCstab\ (\HCustb) destroy stable (unstable) limit cycles as $\kappa$ approaches the homoclinic bifurcation point from above (below).

We come to the following conclusion. In similarity with the case of $M=1$ population, the cases of very small and very strong coupling $\kappa$ result in regimes with quiescent and spiking activity, respectively; both are characterized by high levels of synchrony. In the intermediate regime, the dynamic behavior is more complicated. We thus find the following five stability regions: (i)  for small coupling strength $\kappa$, both populations are quiescent, corresponding to the symmetric configuration \stateQQ\ (white region); (ii) for large coupling strength $\kappa$, both populations are spiking, corresponding to the symmetric configuration \stateSS\ (white region); (iii) for intermediate coupling strengths, we find a region of bistability between the configurations \stateQQ\ and \stateSS\ (gray region); this region of bistability co-exists with asymmetric configurations of either (iv) stationary firing rate, \stateSQ\ or \stateQS\ (red region), or (v) oscillatory firing rates, \stateSQo\ or \stateQSo\ (blue region). In addition, there are regions for intermediate coupling strengths where only \stateQQ\ co-exists stably with \stateSQ\ and \stateQS; or only \stateSS\ with \stateSQ\ and \stateQS\ (see Fig.~\ref{fig:stabilitydiagram}a)).

\section{Discussion\label{sec:discussion}}%
Collective oscillations in neural ensembles are responsible for the rhythm generation required for solving functionally relevant tasks in the brain~\cite{Buzsaki2012,Uhlhaas2006,Marder2001}. Collective oscillations may be facilitated by a variety of network setups, including heterogeneous networks with excitatory and inhibitory coupling leading to gamma rhythms~\cite{Keeley2019,Segneri2020}. Here, we investigated the emergence of collective oscillations in a  simple model consisting of a homogeneous network composed of two (statistically) identical populations of type 1 neurons with non-uniform but symmetric coupling, i.e., neurons are coupled with strength $\kappa$ and $a \kappa $ (with $a\neq 1$) within and between the two populations, respectively.

In this model, each population may assume states corresponding to quiescent (\stateQ) or spiking (\stateS) firing activity.
Thus, we may distinguish symmetric configurations, where both populations are either quiescent or spiking (\stateQQ, \stateSS), and asymmetric configurations, where one population is quiescent but the other is spiking (\stateSQ, \stateQS). We found that stable symmetric configurations may co-exist for certain parameter choices (see Fig.~\ref{fig:bifurcationdiagramOne}a)). We did not find that symmetric configurations are oscillatory except for uniform coupling ($a=1$) or for absent inter-coupling ($a=0$). As we deviate from uniform coupling, $a\neq 1$, \emph{unstable} asymmetric equilibria emerge from symmetric configurations in symmetry-breaking pitchfork bifurcations. Along these solution branches, asymmetric equilibria may further undergo saddle-node bifurcations and thus gain stability (see Fig.~\ref{fig:bifurcationdiagramOne}b)). Asymmetric oscillatory configurations (\stateQSo, \stateSQo) emerge in Hopf bifurcations (Fig.~\ref{fig:bifurcationdiagramTwo}) that  are organized around higher codimension bifurcation points. Depending on parameters, symmetric and asymmetric configurations may be stable and co-exist, resulting in multistability between either stationary configurations only (\stateQQ, \stateSS, \stateQS, \stateSQ); or between stationary and oscillatory configurations (\stateQQ, \stateSS, \stateQSo, \stateSQo). For these regions of stability we have determined valid parameter regions and stability boundaries (Fig.~\ref{fig:stabilitydiagram}).

Oscillator networks with such modular network structure are known to exhibit a high degree of multistability, i.e., depending on initial conditions, a variety of dynamic configurations for the collective states may be assumed in each population. A prominent example are synchronization patterns known as chimera states in Kuramoto oscillator networks~\cite{Abrams2008,Martens2010bistable,PanaggioAbramsReview2015}, which may be employed to store memory or perform computations~\cite{BickMartens2015} or direct the flow of information between populations~\cite{DeschleDaffertshoferBattagliaMartens2019,BickMartens2020}. However, compared to Kuramoto networks with rigidly rotating oscillators, the excitable nature of neurons intrinsically leads to more complicated dynamics and synchronization behavior~\cite{Calugaru2019}. While complicated dynamics may arise in networks composed of identical Kuramoto oscillators  arranged with at least two populations (as well as for broken  parameter symmetries)~\cite{MartensPanaggioAbrams2016,Bick2018}, excitable type 1 neurons produce rich bifurcation behavior and bistability already for a single population, as illustrated in Fig.~\ref{fig:bifurcationdiagramOne} and discussed in~\cite{Luke2013,BickMartens2020}. Such multistability is of great interest in applications, e.g. in neuroscience. A recent study modeled networks of type 1 neurons and demonstrated how the bistability between low and high firing activity --- resulting in a large configuration space that scales with the number of populations --- may be employed to solve cognitive tasks such as memory storage and recall~\cite{Schmidt2018}.

Several studies considered networks of type 1 neurons in terms of their macroscopic behavior. The collective dynamics of a single population was studied in terms of  non-identical Theta neurons with non-zero pulse width~\cite{Luke2013,So2013}, of the response to an external (rigid) forcing~\cite{Luke2014}, of quadratic integrate-and-fire neurons~\cite{Montbrio2015}, and of different coupling functions, oscillations and aging transitions~\cite{Ratas2016}, and of the role of distributed delay in the coupling function~\cite{Ratas2018}.
Luke {\it et al}~\cite{Luke2014} studied a two population model similar to ours; however, they considered unidirectional coupling. This driver-response system exhibits some of the bifurcation structures and collective macroscopic behaviors that we reported here: i.e., the response population exhibits multistable equilibrium states and limit cycles. In addition, their system exhibits chaotic behavior, which was also reported by Ceni {\it et al.}~\cite{Ceni2020} who considered a similar setup, but with exponentially decaying synapses leading to three dimensional dynamics for the macroscopic firing rate equations. Unlike their study, we did not observe quasiperiodic and chaotic dynamics. While we restricted our study to symmetric parameter configurations between the two populations, future research might address the question if breaking parameter symmetries between the populations (such as the coupling strength) may induce bifurcations leading to chaos. For such cases, one may envision torus bifurcations emerging from the Zero-Hopf bifurcations~\cite{Kuznetsov1998}, offering a route to chaos  via bifurcations of Shil'nikov homoclinic orbits to saddle foci.
Ratas and Pyragas~\cite{Ratas2017} studied a network of  quadratic integrate-and-fire neurons with two populations. While their system is similar to ours, it differs in some important aspects. Firstly, neurons are considered to be strongly heterogeneous with an excitability spread around $\hat{\eta} = 0$, thus resulting in a network including both excitable and spiking neurons; here, the majority of neurons are excitable. Secondly, for the coupling function, they use a  threshold modulation coupling function corresponding to a Heaviside function. This system exhibits steady and oscillatory states with symmetric and asymmetric character, but unlike our system, also chaotic behavior and states characterized by anti-phase configurations.

To study collective oscillations of firing activity in our model, it is necessary to deviate from the case of instantaneous pulse coupling ($s\rightarrow\infty$) where collective oscillations are absent (see Appendix~\ref{app:collective_oscillations} and~\cite{Ratas2016,Devalle2017}). The pulse width given by Eq.~\eqref{eq:AriaratnamStrogatzPulseShape} with $s=1$ was large; other pulse shape models~\cite{Montbrio2015,Gerstner2014} may be more realistic and consider that incident pulses arrive instantaneously in order to decay exponentially fast upon arrival over a characteristic time scale $\tau$. It is then frequently assumed that $\tau\rightarrow 0$, resulting in time-symmetric and instantaneous pulses. This strategy certainly simplifies analysis; yet, it appears that this limit biophysically is no more realistic, especially since it results in the same macroscopic equations as given by~\eqref{eq:FREM1} and~\eqref{eq:FREM2} for the limit of $s\rightarrow\infty$;  this again rules out the potential to produce any macroscopic oscillations. For a future study it might be interesting to examine how the specific choice of pulse shape in terms of width and time-asymmetry affects the unfolding of bifurcations.
While many studies either studied small values of $s$ or $s\rightarrow\infty$, it would be interesting to see how the bifurcation scenarios reported in this study translate to the case of causal synaptic potentials that decay exponentially in time ~\cite{Coombes2019next}.

Many questions remain. For instance, breaking parameter symmetry may result in richer dynamics~\cite{Martens2016} including chaos~\cite{Bick2018}; is chaotic motion feasible if excitability parameters ($\hat{\eta}_\sigma$ and $\Delta_\sigma$) are non-identical, or if small delay is introduced in the coupling? Are bifurcation scenarios for spiking neurons ($\hat{\eta}>0$) equally complicated as the ones we observed here for excitable neurons ($\hat{\eta}<0$)? In terms of switching between configurations and devising a control method to do this, it may be useful to determine basins of attraction for the various configurations or responses to directed perturbations~\cite{MartensPanaggioAbrams2016,Schmidt2018}. Furthermore, networks with larger population number $M>2$ provide a larger set of dynamic configurations~\cite{Shanahan2010,Wildie2012,Schmidt2018}; but how large is the set of configurations as a function of the population number, and which of the configurations are stable, and which oscillatory? Future studies may address such and further questions.

\section*{Acknowledgements}%
The authors would like to thank C.~Bick  and B.~Pietras for helpful discussions, and A.~Torcini and P.~So for helpful correspondence.
Research conducted by B.J. is partially supported by funding from EU-COST Technical University of Denmark.

\section*{AIP Data Sharing Policy}
Data sharing is not applicable to this article as no new data were created or analyzed in this study.

\appendix

\section{Collective oscillations for non-zero pulse width ($s<\infty$) \label{app:collective_oscillations}}
We briefly discuss the existence of Hopf bifurcations and resulting limit cycle oscillations in the firing rate $r(t)$ for varying pulse shape parameter $ s $, for the simple case of $M=1$ population. To determine the presence of Hopf instabilities, we examine eigenvalues of the Jacobian of \eqref{eq:FREM1},
\begin{align}
 J &= \left(
 \begin{array}{cc}
  2v & 2r\\
  -2\pi^2 r  + \frac{\partial }{\partial r} I &   2v +\frac{\partial }{\partial v} I\
 \end{array}
 \right) .\
\end{align}
Steady state of \eqref{eq:FREM1} implies $v^*=-\frac{\Delta}{2\pi}\frac{1}{r^*}$ so that
\begin{align}
 \text{tr}{(J)}
  &= -\frac{2\Delta}{\pi}\frac{1}{r^*} +\frac{\partial }{\partial  v} I|_{(r,v)=(r^*,v^*)}.\
\end{align}

A necessary condition for a Hopf bifurcation is that $ \text{tr}{(J)}=0 $.
For the case of infinitely narrow pulses, $ s = \infty $, Hopf bifurcations are impossible: we have $ \frac{\partial }{\partial v} I = \kappa \frac{\partial }{\partial v} P^{(\infty)} = 0 $ and thus $\text{tr}{(J)} = -2\Delta/\pi /r^* <0 $ for all $ r^*>0 $. Hence, Hopf bifurcations and resulting limit cycles regardless of the choice of parameters can be ruled out for this case.

Conversely, we know that Hopf bifurcations are possible for $s=1$ (see Fig.~\ref{fig:bifdiag_M1}) and $s=2$ (see Luke {\it et al.}~\cite{Luke2013}).
Indeed, the trace for $ 1 < s < \infty $ involves more complicated terms, and Hopf bifurcation cannot easily be ruled out. While an analytical proof remains elusive, using a numerical analysis based on solving the zero trace condition, one finds that limit cycle oscillations are feasible for a large range of pulse shape parameters, including at least $ 1 \leq s\leq 20 $.

\section{Methodology \label{app:meth}}
The data for the bifurcation diagrams in Figs.~\ref{fig:bifurcationdiagramOne} and~\ref{fig:bifurcationdiagramTwo} was obtained via numerical continuation of equilibria and limit cycles (using MatCont software) in the parameter $\kappa$; thus we encountered codimension-1 bifurcation points \SNone, \SNtwo, \SNthree, \SNfour, \HBstab, \HBustb, \HBustbstab, \HBprime, \HCstab, \HCustb, \HCprime, \SNLCone, \SNLCtwo, \PFone, \PFtwo. With the exception of \PFone\ and \PFtwo\ we continued all these degenerate states as bifurcation curves in the parameters $\kappa$ and $a$ using MatCont (Fig.~\ref{fig:stabilitydiagram}); thereby we detected the codimension $\geq2$ bifurcation points reported in Sec.~IV~3.
The direct two-parameter continuation of the bifurcation curves \PFone, \PFtwo\  posed technical problems when using MatCont; instead we therefore determined the loci of \PFone\ and \PFtwo\ by computing bifurcation diagrams in a single parameter, $\kappa$, for set values of $a$, resulting in the parameter list $(\kappa,a)$ in Tab.~\ref{tab:PFpoints}. The curves shown in Fig.~\ref{fig:stabilitydiagram} (dashed black curves) are splines interpolating these data points.
\begin{table}
\begin{tabular}{c|c|c}
$a$ & $\kappa$ (\PFone) & $\kappa$ (\PFtwo)\\
\hline
0.7& 1.476& 5.546\\
0.65& 1.438& 5.728\\
0.6& 1.414& 5.915\\
0.5& 1.400& 6.320\\
0.4& 1.419& 6.777\\
0.35& 1.439& 7.029\\
0.25& 1.500& 7.594\\
0.204& 1.538& 7.884\\
0.18& 1.561& 8.045\\
0.1& 1.652& 8.630\\
-0.01& 1.824& 9.590\\
-0.05& 1.904& 9.993\\
-0.1& 2.020& 10.548\\
-0.15& 2.160& 11.169\\
-0.2& 2.329& 11.867\\
-0.27& 2.632& 13.004\\
-0.35&3.117&14.604\\
-0.4&3.538&15.821\\
\hline
\end{tabular}
\caption{\label{tab:PFpoints}Bifurcation points numerically detected for \PFone\ and \PFtwo.}
\end{table}

\bibliographystyle{unsrt}

\begin{thebibliography}{10}

\bibitem{Ermentrout1986}
G.~B. Ermentrout and N.~Kopell.
\newblock {Parabolic Bursting in an excitable system coupled with a slow
  oscillation}.
\newblock {\em SIAM Journal on Applied Mathematics}, 46(2):233--253, 1986.

\bibitem{Latham2000}
P.~E. Latham, B.~J. Richmond, P.~G. Nelson, and S.~Nirenberg.
\newblock {Intrinsic dynamics in neuronal networks. I. Theory}.
\newblock {\em Journal of Neurophysiology}, 83(2):808--827, 2000.

\bibitem{Hansel2001}
D.~Hansel and G.~Mato.
\newblock {Existence and stability of persistent states in large neuronal
  networks}.
\newblock {\em Physical Review Letters}, 86(18):4175--4178, 2001.

\bibitem{Gerstner2014}
W.~Gerstner, W.~M. Kistler, R.~Naud, and L.~Paninski.
\newblock {\em {Neuronal dynamics: From single neurons to networks and models
  of cognition}}.
\newblock Cambridge University Press, 2014.

\bibitem{OttAntonsen2008}
E.~Ott and T.~M. Antonsen.
\newblock {Low dimensional behavior of large systems of globally coupled
  oscillators}.
\newblock {\em Chaos (Woodbury, N.Y.)}, 18(3):037113, sep 2008.

\bibitem{Montbrio2015}
E.~Montbri{\'o}, D.~Paz{\'o}, and A.~Roxin.
\newblock {Macroscopic description for networks of spiking neurons}.
\newblock {\em Physical Review X}, 5(2):1--15, 2015.

\bibitem{Luke2013}
T.~B. Luke, E.~Barreto, and P.~So.
\newblock {Complete Classification of the Macroscopic Behavior of a
  Heterogeneous Network of Theta Neurons}.
\newblock {\em Neural Computation}, 25:3207--3234, 2013.

\bibitem{BickMartens2020}
C.~Bick, C.~Laing, M.~Goodfellow, and E.A. Martens.
\newblock {Understanding the dynamics of biological and neural oscillator
  networks through exact mean-field reductions: a review}.
\newblock {\em Journal of Mathematical Neuroscience}, 9(10), 2020.

\bibitem{Byrne2019}
Á. Byrne, D.~Avitabile, and S.~Coombes.
\newblock {Next-generation neural field model: The evolution of synchrony
  within patterns and waves}.
\newblock {\em Physical Review E}, 2019.

\bibitem{Marder2001}
E.~Marder and D.~Bucher.
\newblock {Central pattern generators and the control of rythmic movements}.
\newblock {\em Current Biology}, 11:R986--R996, 2001.

\bibitem{Smith1991}
J.~C. Smith, H.~H. Ellenberger, K.~Ballanyi, D.~W. Richter, and J.~L. Feldman.
\newblock {Pre-B{\"o}tzinger Complex: A Brainstem Region That May Generate
  Respiratory Rhythm in Mammals.}
\newblock {\em Science}, 254(5032):726--729, 1991.

\bibitem{Buzsaki2012a}
G.~Buzs{\'a}ki and X.-J. Wang.
\newblock {Mechanisms of Gamma Oscillations}.
\newblock {\em Annual Review of Neuroscience}, 35(1):203--225, 2012.

\bibitem{Bullmore2009}
E.~T. Bullmore and O.~Sporns.
\newblock {Complex brain networks: graph theoretical analysis of structural and
  functional systems.}
\newblock {\em Nature reviews. Neuroscience}, 10(3):186--98, 2009.

\bibitem{Meunier2010}
D.~Meunier, R.~Lambiotte, and E.~T. Bullmore.
\newblock {Modular and hierarchically modular organization of brain networks}.
\newblock {\em Frontiers in Neuroscience}, 4(DEC):1--11, 2010.

\bibitem{Harris2005}
K.~D. Harris.
\newblock {Neural signatures of cell assembly organization : Article : Nature
  Reviews Neuroscience}.
\newblock {\em Nature Reviews}, 6(May):399--407, 2005.

\bibitem{Lynn2019}
C.~W. Lynn and D.~S. Bassett.
\newblock {The physics of brain network structure, function and control}.
\newblock {\em Nature Reviews Physics}, 2019.

\bibitem{Glass2001}
L.~Glass.
\newblock {Synchronization and rhythmic processes in physiology}.
\newblock {\em Nature}, 410(March):277--284, 2001.

\bibitem{Uhlhaas2006}
P.~J. Uhlhaas and W.~Singer.
\newblock {Neural synchrony in brain disorders: relevance for cognitive
  dysfunctions and pathophysiology.}
\newblock {\em Neuron}, 52(1):155--68, 2006.

\bibitem{Fell2011}
J.~Fell and N.~Axmacher.
\newblock {The role of phase synchronization in memory processes}.
\newblock {\em Nature Reviews Neuroscience}, 12(2):105--118, 2011.

\bibitem{Fries2009}
P.~Fries.
\newblock {Neuronal gamma-band synchronization as a fundamental process in
  cortical computation}.
\newblock {\em Annual Review of Neuroscience}, 32:209--24, 2009.

\bibitem{Wang2010}
X.~J. Wang.
\newblock {Neurophysiological and Computational Principles of Cortical Rhythms
  in Cognition}.
\newblock {\em Physiological Reviews}, 90(3):1195--1268, 2010.

\bibitem{Singer1995}
W.~Singer and C.~M. Gray.
\newblock {Visual feature integration and the temporal correlation hypothesis}.
\newblock {\em Ann Rev Neurosci}, 18:555--586, 1995.

\bibitem{Fries2005}
P.~Fries.
\newblock {A mechanism for cognitive dynamics: neuronal communication through
  neuronal coherence.}
\newblock {\em Trends in Cognitive Sciences}, 9(10):474--80, 2005.

\bibitem{Rabinovich2012}
M.~I. Rabinovich, V.~S. Afraimovich, C.~Bick, and P.~Varona.
\newblock {Information flow dynamics in the brain}.
\newblock {\em Physics of Life Reviews}, 9(1):51--73, 2012.

\bibitem{Kirst2016}
C.~Kirst, M.~Timme, and D.~Battaglia.
\newblock {Dynamic information routing in complex networks}.
\newblock {\em Nature Communications}, 7:11061, 2016.

\bibitem{DeschleDaffertshoferBattagliaMartens2019}
N.~Deschle, A.~Daffertshofer, D.~Battaglia, and E.A. Martens.
\newblock {Directed Flow of Information in Chimera States}.
\newblock {\em {Frontiers in Applied Mathematics and Statistics}}, 5(28),
  {2019}.

\bibitem{Abrams2008}
D.~M. Abrams, R.~Mirollo, S.~H. Strogatz, and D.~A. Wiley.
\newblock {Solvable model for chimera states of coupled oscillators}.
\newblock {\em Phys. Rev. Lett}, 101:084103, 2008.

\bibitem{MartensPanaggioAbrams2016}
E.~A. Martens, M.~J. Panaggio, and D.~M. Abrams.
\newblock {Basins of Attraction for Chimera States}.
\newblock {\em New Journal of Physics, Fast Track Communication}, 18:022002,
  2016.

\bibitem{Martens2010bistable}
E.~A. Martens.
\newblock {Bistable Chimera Attractors on a Triangular Network of Oscillator
  Populations}.
\newblock {\em Physical Review E}, 82(1):016216, jul 2010.

\bibitem{Martens2016}
E.~A. Martens, C.~Bick, and M.~J. Panaggio.
\newblock {Chimera states in two populations with heterogeneous phase-lag}.
\newblock {\em Chaos}, 26(9):094819, 2016.

\bibitem{Laing2009}
C.~R. Laing.
\newblock {The dynamics of chimera states in heterogeneous Kuramoto networks}.
\newblock {\em Physica D: Nonlinear Phenomena}, 238(16):1569--1588, aug 2009.

\bibitem{Laing2012a}
C.~R. Laing, K.~Rajendran, and I.~G. Kevrekidis.
\newblock {Chimeras in random non-complete networks of phase oscillators}.
\newblock {\em Chaos: An Interdisciplinary Journal of Nonlinear Science},
  22(1):013132, 2012.

\bibitem{Scholl2016}
E.~Sch{\"o}ll.
\newblock {Synchronization patterns and chimera states in complex networks:
  Interplay of topology and dynamics}.
\newblock {\em The European Physical Journal Special Topics},
  225(6-7):891--919, 2016.

\bibitem{PanaggioAbramsReview2015}
M.~J. Panaggio and D.~M. Abrams.
\newblock {Chimera states: coexistence of coherence and incoherence in networks
  of coupled oscillators}.
\newblock {\em Nonlinearity}, 28(3):R67, 2015.

\bibitem{Laing2016}
C.~R. Laing.
\newblock {Phase oscillator network models of brain dynamics}.
\newblock In Ahmed~A. Moustafa, editor, {\em {Computational Models of Brain and
  Behavior}}, chapter~37, pages 505--518. Wiley-Blackwell, 2017.

\bibitem{Luke2014}
T.~B. Luke, E.~Barreto, and P.~So.
\newblock {Macroscopic complexity from an autonomous network of networks of
  theta neurons}.
\newblock {\em Frontiers in Computational Neuroscience}, 8(NOV):1--11, 2014.

\bibitem{Schmidt2018}
H.~Schmidt, D.~Avitabile, E.~Montbri{\'o}, and A.~Roxin.
\newblock {Network mechanisms underlying the role of oscillations in cognitive
  tasks}.
\newblock {\em PLoS Computational Biology}, 14(9):1--24, 2018.

\bibitem{YamakouHjorthMartens2020}
M.~Yamakou, P.~G. Hjorth, and E.~A. Martens.
\newblock {Optimal self-induced stochastic resonance in multiplex neural
  networks: electrical versus chemical synapses}.
\newblock {\em {Frontiers in Neuroscience}}, 14(62), {2020}.

\bibitem{Weerasinghe2018}
G.~Weerasinghe, B.~Duchet, H.~Cagnan, P.r Brown, C.~Bick, and R.~Bogacz.
\newblock {Predicting the effects of deep brain stimulation using a reduced
  coupled oscillator model}.
\newblock {\em PLoS Comp. Biology}, 15(8), 2018.

\bibitem{Watanabe1994}
S.~Watanabe and S.~H. Strogatz.
\newblock {Constants of motion for superconducting Josephson arrays}.
\newblock {\em Physica D}, 74(3-4):197--253, 1994.

\bibitem{Laing2018}
C.~R. Laing.
\newblock {The Dynamics of Networks of Identical Theta Neurons}.
\newblock {\em Journal of Mathematical Neuroscience}, 8(1), 2018.

\bibitem{Wilson1972}
H.~R. Wilson and J.~D. Cowan.
\newblock {Excitatory and inhibitory interactions in localized populations of
  model neurons}.
\newblock {\em Biophysical journal}, 12(1):1--24, 1972.

\bibitem{Amari1977}
S.~Amari.
\newblock {Dynamics of pattern formation in lateral-inhibition type neural
  fields}.
\newblock {\em Biological Cybernetics}, 27(2):77--87, 1977.

\bibitem{Lin2020}
L.~Lin, E.~Barreto, and P.~So.
\newblock {Synaptic Diversity Suppresses Complex Collective Behavior in
  Networks of Theta Neurons}.
\newblock {\em {Frontiers in Computational Neuroscience}}, 14(May):1--17, 2020.

\bibitem{Buzsaki2012}
G.~Buzs{\'a}ki and B.~O. Watson.
\newblock {Brain rhythms and neural syntax: Implications for efficient coding
  of cognitive content and neuropsychiatric disease}.
\newblock {\em Dialogues in Clinical Neuroscience}, 2012.

\bibitem{Fries2007gamma}
P.~Fries, D.~Nikoli{\'c}, and W.~Singer.
\newblock {The gamma cycle}.
\newblock {\em Trends in neurosciences}, 30(7):309--316, 2007.

\bibitem{Keeley2017}
S.~Keeley, A.~Fenton, and J.~Rinzel.
\newblock {Modeling fast and slow gamma oscillations with interneurons of
  different subtype}.
\newblock {\em Journal of Neurophysiology}, 117(3):950--965, 2017.

\bibitem{Keeley2019}
S.~Keeley, A.~Byrne, A.~Fenton, and J.~Rinzel.
\newblock {Firing rate models for gamma oscillations}.
\newblock {\em Journal of Neurophysiology}, 121(6):2181--2190, 2019.

\bibitem{Segneri2020}
M.~Segneri, H.~Bi, S.~Olmi, and A.~Torcini.
\newblock {Theta-nested gamma oscillations in next generation neural mass
  models}.
\newblock {\em {Frontiers in Computational Neuroscience}}, pages 1--28, 2020.

\bibitem{Devalle2017}
F.~Devalle, A.~Roxin, and E.~Montbri{\'o}.
\newblock {Firing rate equations require a spike synchrony mechanism to
  correctly describe fast oscillations in inhibitory networks}.
\newblock {\em PLoS Computational Biology}, 13(12):1--21, 2017.

\bibitem{So2013}
P.~So, T.~B. Luke, and E.~Barreto.
\newblock {Networks of theta neurons with time-varying excitability:
  Macroscopic chaos, multistability, and final-state uncertainty}.
\newblock {\em Physica D: Nonlinear Phenomena}, 267:16--26, may 2013.

\bibitem{Dhooge2008}
A.~Dhooge, W.~Govaerts, Yu~A. Kuznetsov, H.~G.E. Meijer, and B.~Sautois.
\newblock {New features of the software MatCont for bifurcation analysis of
  dynamical systems}.
\newblock {\em Mathematical and Computer Modelling of Dynamical Systems}, 2008.

\bibitem{Ermentrout2008}
B.~Ermentrout.
\newblock {Ermentrout-Kopell canonical model}.
\newblock {\em Scholarpedia}, 3(3):1398, 2008.

\bibitem{ariaratnam2001}
J.~T. Ariaratnam and S.~H. Strogatz.
\newblock {Phase diagram for the Winfree model of coupled nonlinear
  oscillators}.
\newblock {\em Physical Review Letters}, 86(19):4278, 2001.

\bibitem{Ermentrout2010}
G.~B. Ermentrout and D.~H. Terman.
\newblock {\em {Mathematical Foundations of Neuroscience}}, volume~35 of {\em
  {Interdisciplinary Applied Mathematics}}.
\newblock Springer, New York, NY, 2010.

\bibitem{Brunel2003}
N.~Brunel and P.~E. Latham.
\newblock {Firing rate of the noisy quadratic integrate-and-fire neuron}.
\newblock {\em Neural Computation}, 15(10):2281--2306, 2003.

\bibitem{Ratas2016}
I.~Ratas and K.~Pyragas.
\newblock {Macroscopic self-oscillations and aging transition in a network of
  synaptically coupled quadratic integrate-and-fire neurons}.
\newblock {\em Physical Review E}, 94(3):1--11, 2016.

\bibitem{Kuznetsov1998}
Y.~A. Kuznetsov.
\newblock {\em {Elements of applied bifurcation theory}}.
\newblock Springer-Verlag, New York, 1998.

\bibitem{Guckenheimer2007GH}
J.~Guckenheimer and Y.~Kuznetsov.
\newblock {Bautin bifurcation}.
\newblock {\em Scholarpedia}, 2007.

\bibitem{Guckenheimer2007ZH}
J.~Guckenheimer and Y.~Kuznetsov.
\newblock {Fold-Hopf bifurcation}.
\newblock {\em Scholarpedia}, 2007.

\bibitem{BickMartens2015}
C.~Bick and E.~A. Martens.
\newblock {Controlling Chimeras}.
\newblock {\em New Journal of Physics}, 17:033030, 2015.

\bibitem{Calugaru2019}
D.~C{\u a}lug{\u a}ru, J.~F. Totz, E.~A. Martens, and H.~Engel.
\newblock {First-order synchronization transition in a large population of
  strongly coupled relaxation oscillators}.
\newblock {\em Science Advances}, {6}(39):eabb2637, {2020}.

\bibitem{Bick2018}
C.~Bick, M.~J. Panaggio, and E.~A. Martens.
\newblock {Chaos in Kuramoto Oscillator Networks}.
\newblock {\em Chaos}, 28:071102, 2018.

\bibitem{Ratas2018}
I.~Ratas and K.~Pyragas.
\newblock {Macroscopic oscillations of a quadratic integrate-and-fire neuron
  network with global distributed-delay coupling}.
\newblock {\em Physical Review E}, 98(5):1--11, 2018.

\bibitem{Ceni2020}
A.~Ceni, S.~Olmi, A.~Torcini, and D.~Angulo-Garcia.
\newblock {Cross frequency coupling in next generation inhibitory neural mass
  models}.
\newblock {\em Chaos (Woodbury, N.Y.)}, 30(5):053121, 2020.

\bibitem{Ratas2017}
I.~Ratas and K.~Pyragas.
\newblock {Symmetry breaking in two interacting populations of quadratic
  integrate-and-fire neurons}.
\newblock {\em Physical Review E}, 96(4):1--9, 2017.

\bibitem{Coombes2019next}
Stephen Coombes and {\'A}ine Byrne.
\newblock Next generation neural mass models.
\newblock In {\em Nonlinear Dynamics in Computational Neuroscience}, pages
  1--16. Springer, 2019.

\bibitem{Shanahan2010}
M.~Shanahan.
\newblock {Metastable chimera states in community-structured oscillator
  networks.}
\newblock {\em Chaos (Woodbury, N.Y.)}, 20(1):013108, mar 2010.

\bibitem{Wildie2012}
M.~Wildie and M.~Shanahan.
\newblock {Metastability and chimera states in modular delay and pulse-coupled
  oscillator networks}.
\newblock {\em Chaos: An Interdisciplinary Journal of Nonlinear Science},
  22(4):043131, 2012.

\end{thebibliography}

\end{document}